\documentclass{article}
\usepackage{titletoc} 
\usepackage{arxiv}
\usepackage{appendix}
\usepackage[utf8]{inputenc} 
\usepackage[T1]{fontenc}    
\usepackage{hyperref}       
\hypersetup{
    colorlinks = true,      
    linkcolor  = black,     
    urlcolor   = blue       
}
\usepackage{url}            
\usepackage{booktabs}       
\usepackage{amsfonts}       
\usepackage{nicefrac}       
\usepackage{microtype}      
\usepackage{lipsum}		
\usepackage{graphicx}
\usepackage{doi}
\usepackage{float}  
\usepackage{caption} 
\usepackage{amsmath}
\usepackage{array}      
\usepackage{subcaption} 
\usepackage{geometry}   
\usepackage{multirow}   
\usepackage{booktabs}   
\usepackage{caption}    

\title{Deconstructing Mobility Segregation: A Network Analysis of Racialized Flows in Pandemic-Era NYC}

\author{ \includegraphics[scale=0.06]{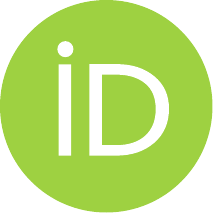}\hspace{1mm}Wei-Peng Nie\\
	Hebei Key Laboratory of Future Urban Intelligent Traffic Management\\ Beijing Jiaotong University\\
	Beijing 100044, Chian \\
	\texttt{wpnie@bjtu.edu.cn} \\
    \And
    \includegraphics[scale=0.06]{orcid.pdf}\hspace{1mm}Tian-Rong Ding\\
	Department of Industrial Engineering\\
    Tsinghua University, \\
	Beijing 100044, Chian \\
	\texttt{dtr@mail.tsinghua.edu.cn} \\
	\And
	\includegraphics[scale=0.06]{orcid.pdf}\hspace{1mm}Xiao-Yong Yan* \\
	Hebei Key Laboratory of Future Urban Intelligent Traffic Management\\ Beijing Jiaotong University\\
	Beijing 100044, Chian \\
	\texttt{yanxy@bjtu.edu.cn} \\
    \And
	\includegraphics[scale=0.06]{orcid.pdf}\hspace{1mm}Tao Zhou* \\
	Comple$\chi$ Lab, Big Data Research Center
\\
	University of Electronic Science and Technology of China\\
	Chengdu 610054, China \\
	\texttt{zhutou@ustc.edu.cn} \\
    \And
	\includegraphics[scale=0.06]{orcid.pdf}\hspace{1mm}Zi-You Gao* \\
	Hebei Key Laboratory of Future Urban Intelligent Traffic Management\\ Beijing Jiaotong University\\
	Beijing 100044, Chian \\
	\texttt{zygao@bjtu.edu.cn} \\
}

\renewcommand{\shorttitle}{\textit{arXiv} Template}

\hypersetup{
pdftitle={The Anatomy of Mobility Segregation: Racialized Flow Networks in NYC During COVID-19},
pdfauthor={Wei-Peng Nie},
pdfkeywords={Racial Segregation, Mixing Pattern, Active Homophily, Pandemic Mobility, Gravity Model},
}

\begin{document}

\maketitle

\begin{abstract}
Urban segregation research has long relied on residential patterns, yet growing evidence suggests that racial/ethnic segregation also manifests systematically in mobility behaviors. Leveraging anonymized mobile device data from New York City before and during the COVID-19 pandemic, we develop a network-analytic framework to dissect mobility segregation in racialized flow networks. We examine citywide racial mixing patterns through mixing matrices and assortativity indices, revealing persistent diagonal dominance where intra-group flows constituted $69.27\%$ of total movements. Crucially, we develop a novel dual-metric framework that reconceptualizes mobility segregation as two interlocking dimensions: structural segregation—passive exposure patterns driven by residential clustering, and preferential segregation—active homophily in mobility choices beyond spatial constraints. Our gravity-adjusted indices reveal that racial divisions transcend residential clustering---all racial groups exhibit active homophily after spatial adjustments, with particularly severe isolation during the pandemic. Findings highlight how pandemic restrictions disproportionately amplified asymmetric isolation, with minority communities experiencing steeper declines in access to White-dominated spaces. Finally, we propose a Homophily Gravity Model with racial similarity parameter, which significantly improves the prediction accuracy of race-specific mobility flows and more accurately reproduces intra-group mobility preferences and directional exposure patterns.
Overall, this study redefines mobility segregation as a multidimensional phenomenon, where structural constraints, active preferences, and crisis responses compound to reshape urban racial inequality.

\end{abstract}


\section{Introduction}

Cities have long served as hubs of economic opportunity and cultural exchange, attracting diverse populations whose interactions shape the social fabric of urban life \cite{balland2020complex,liu2022revealing}. However, this diversity often coexists with persistent segregation along racial and socioeconomic lines, particularly in the United States, where historical policies like redlining and discriminatory lending have entrenched spatial inequalities \cite{florida2017new, abramson1995changing}. Segregation not only limits access to resources like employment opportunities, quality schools and healthcare but also reinforces stereotypes, reduces cross-group trust, and perpetuates cycles of disadvantage \cite{bor2017population, bowen1995toward}. In the U.S. context, racial segregation has been especially pronounced between White and non-White communities, with Black, Hispanic and Asian populations frequently confined to neighborhoods with fewer opportunities—a pattern that extends to their daily mobility and access to urban amenities \cite{massey1990american,park2019city,torrats2021using}. 

Traditional approaches in research on segregation relied on census data to measure the uneven distribution of racial groups across neighborhoods, using indices like dissimilarity or isolation index \cite{feitosa2007global, browning2017socioeconomic}. Recent research has increasingly demonstrated that residential segregation patterns provide an incomplete picture of urban socio-spatial divisions \cite{kwan2013beyond, candipan2021residence}. A more comprehensive understanding requires examining segregation dynamics across the full spectrum of daily activities \cite{wong2011measuring, xu2024experienced}. With the advent of large-scale mobility data, contemporary work leverages mobile phone records, GPS traces, and social media data to examine the probability of visits and contacts of residents to activity spaces across work, leisure, and transit locations \cite{xu2019quantifying, silm2014temporal, li2023quantifying,hilman2022socioeconomic}. These studies reveal that mobility segregation manifests differently across activity types - sometimes stronger, sometimes weaker than residential patterns - highlighting how daily mobility both reinforces and mitigates urban divisions \cite{liao2025socio}. The emerging "experienced segregation" paradigm quantifies actual cross-group exposure through co-presence metrics in shared spaces \cite{hilman2023mobility,athey2021estimating, moro2021mobility}, while activity space analyses track segregation across individuals' complete mobility networks \cite{park2018beyond,wu2023revealing}. These studies have demonstrated that segregation is not static but actively reproduced through daily mobility choices, with implications for access to jobs, services, and social networks \cite{hu2019segregation,xu2025using}.

Despite these advances, critical gaps remain. First, while existing studies quantify segregation through point-to-point visitation probabilities or co-presence metrics at individual locations \cite{moro2021mobility,park2018beyond}, they neglect how the emergent structure of racialized flow networks reflects the segregation dynamics at the city level. Critically, these structural properties remain invisible to location-scale analyses, as they only manifest when examining the full mobility network as an interconnected system. Second, current approaches predominantly treat mobility segregation as a monolithic outcome of demographic disparities, failing to distinguish its constitutive mechanisms that operate through distinct behavioral channels. Therefore, existing mobility-based metrics struggle to disentangle whether segregation arises from passive exposure (e.g., increased co-racial encounters attributable solely to residential clustering patterns) or from active homophily (e.g., individuals avoiding cross-racial destinations). This distinction is crucial: interventions targeting housing disparities will differ from those addressing behavioral barriers to integration \cite{chetty2014land,logan2025racial}. Without clarifying these mechanisms, policies risk addressing symptoms rather than root causes of urban division. Third, little is known about how large-scale shocks like pandemics reshape these systemic network interactions and amplify or transform active mobility preferences - a crucial oversight given the persistent urban crises cities face \cite{yabe2025behaviour}.

To address these critical gaps, we employ network science approaches to analyze urban mobility segregation in racialized mobility flows and examine the evolutionary dynamics under pandemic impact. Leveraging anonymized mobile device data in New York City during COVID-19, we construct the racialized mobility network and reconceptualizes urban segregation as systemic imbalances in inter-group connectivity. Our study first examines citywide racial mixing patterns through mobility mixing matrices and assortativity indices, revealing persistent diagonal dominance where intra-group flows constituted the predominant component. To disentangle structural (passive exposure) and preferential segregation (active homophily), we develop a dual-metric analytical framework: (1) a baseline null network accounting only for tract-level racial demographics, and (2) an enhanced gravity-based null model incorporating both demographic distributions and spatial distances. The residual flows exceeding these null expectations --- quantified through our novel segregation indices --- isolate active homogeneous mobility, with results demonstrating that significant preferential segregation persists in mobility patterns and uncovering the potentially serious isolation situation within Asian communities. Crucially, multi-dimensional segregation analysis reveals that mobility segregation systematically intensified during the pandemic, revealing that containment measures failed to adequately account for pre-existing segregation patterns. We further formalize how racial preferences modulate mobility networks and propose the Homophily Gravity Model containing a racial similarity parameter, which improves flow prediction accuracy compared to baseline specifications. Overall, the simultaneous measurement of these interlocking dimensions captures the compounding nature of segregation dynamics and provides a comprehensive assessment of how crisis conditions reshape both the structural and behavioral foundations of urban mobility segregation.

\section{Result}

\begin{figure}[t]
    \centering 
    \includegraphics[width=13cm,trim=10 0 0 5,clip]{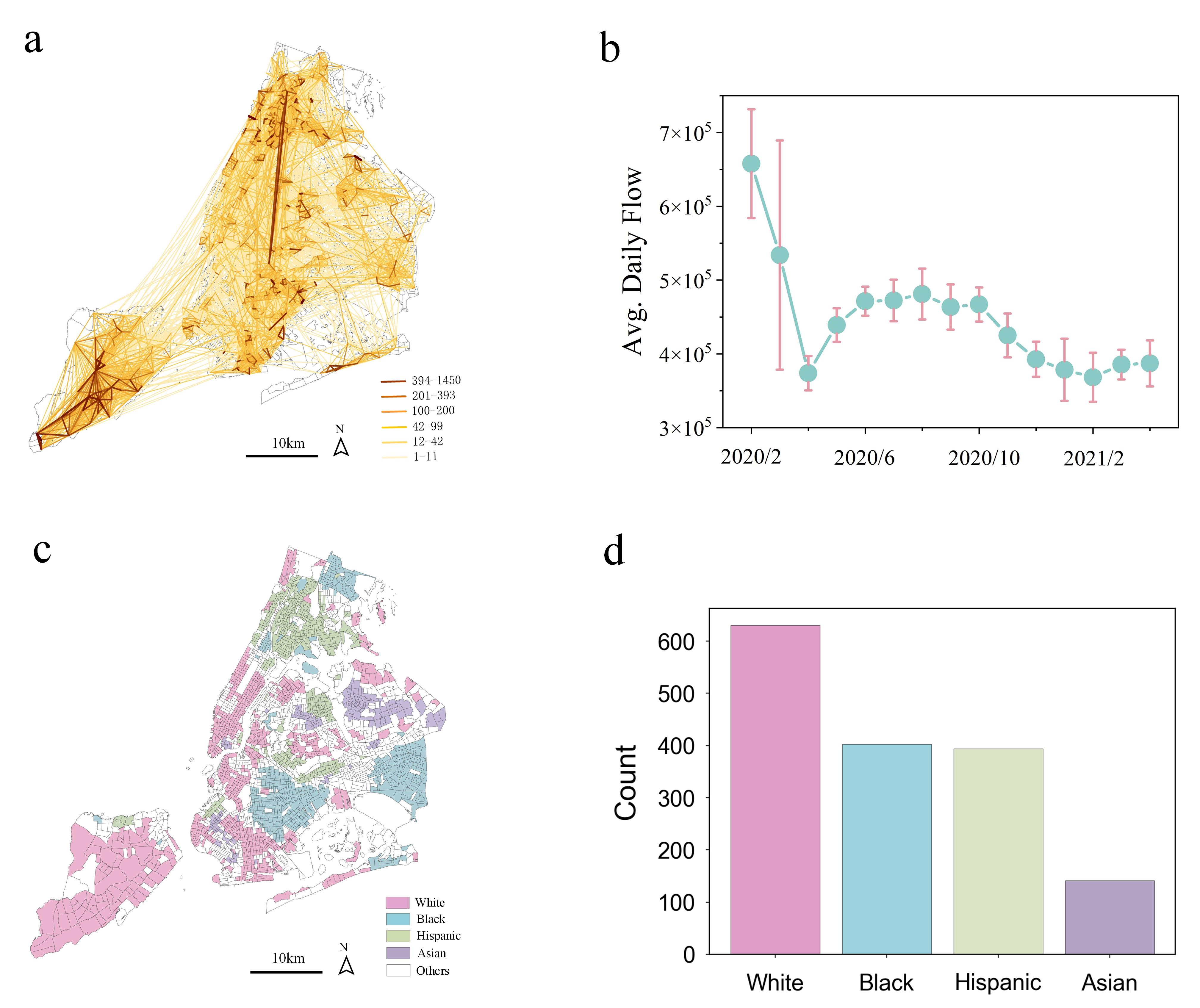}
    \caption{\textbf{a} A visualization figure of daily mobility flows between census tracts (CTs) in New York City on January 1, 2019. \textbf{b} Point plot showing monthly average daily flows (points) with error bars representing within-month variability ($\pm1$ standard deviation). After the implementation of epidemic containment
measures, the daily flow of people dropped rapidly and fluctuated with changes in the measures. \textbf{c} Geographic distribution of census tracts (CTs) dominated by four major racial groups: White (red), Black (blue), Hispanic (green), and Asian (purple).
\textbf{d} Count distribution of majority-group CTs: White (631), Black (403), Hispanic (394), and Asian (141).}
\vspace{-10pt}  
	\label{fig:fig1}
\end{figure}

Our analysis leverages two foundational datasets to examine racial segregation dynamics in New York City during the COVID-19 pandemic. First, we utilize anonymized mobile device location records spanning January 2019 to April 2021 (see Methods section and Supplementary Note 1.1)\cite{kang2020multiscale}, which capture daily visitor flows between census tracts (CTs). \hyperref[fig:fig1]{Fig.1a} shows a visualization figure of daily mobility flows on January 1, 2019. These data obtained approximately 24.4 million anonymized visitor trips monthly during the pre-pandemic baseline period (January 2019-February 2020), enabling precise measurement of mobility pattern shifts across subsequent pandemic phases. Details of containment measures in New York City during the COVID-19 pandemic can be found in Supplementary Note 2. \hyperref[fig:fig1]{Fig.1b} shows monthly average daily flows varying with time, a sharp decline occurred in March 2020 when lockdown measures were implemented, with exceptionally high variability reflecting the transitional period of policy adoption. Flow volumes remained depressed through 2020 before a gradual recovery in 2021. Second, we integrate 2020 Decennial Census data to classify tracts by racial majority and select CTs where a single racial group constitutes $\geq 50\%$ of the population (a threshold of $70\%$ produced similar results, see Supplementary Note 7.2). We focus on the four largest racial groups: White (631 tracts), Black (403 tracts), Hispanic (394 tracts), and Asian (141 tracts), and four groups all show spatial clustering (see \hyperref[fig:fig1]{ Fig.1c} and \hyperref[fig:fig1]{d}). This selection of $1,569$ tracts ($67.4\%$ of NYC’s $2,327$ tracts) excludes $758$ racially mixed tracts to minimize demographic noise; flows associated with racially mixed tracts were also excluded from our analyses. Details of tract selection can be found in Supplementary Note 1.3.
 
\begin{figure}[t!] 
    \centering 
    \includegraphics[width=16cm,trim=0 0 10 0,clip]{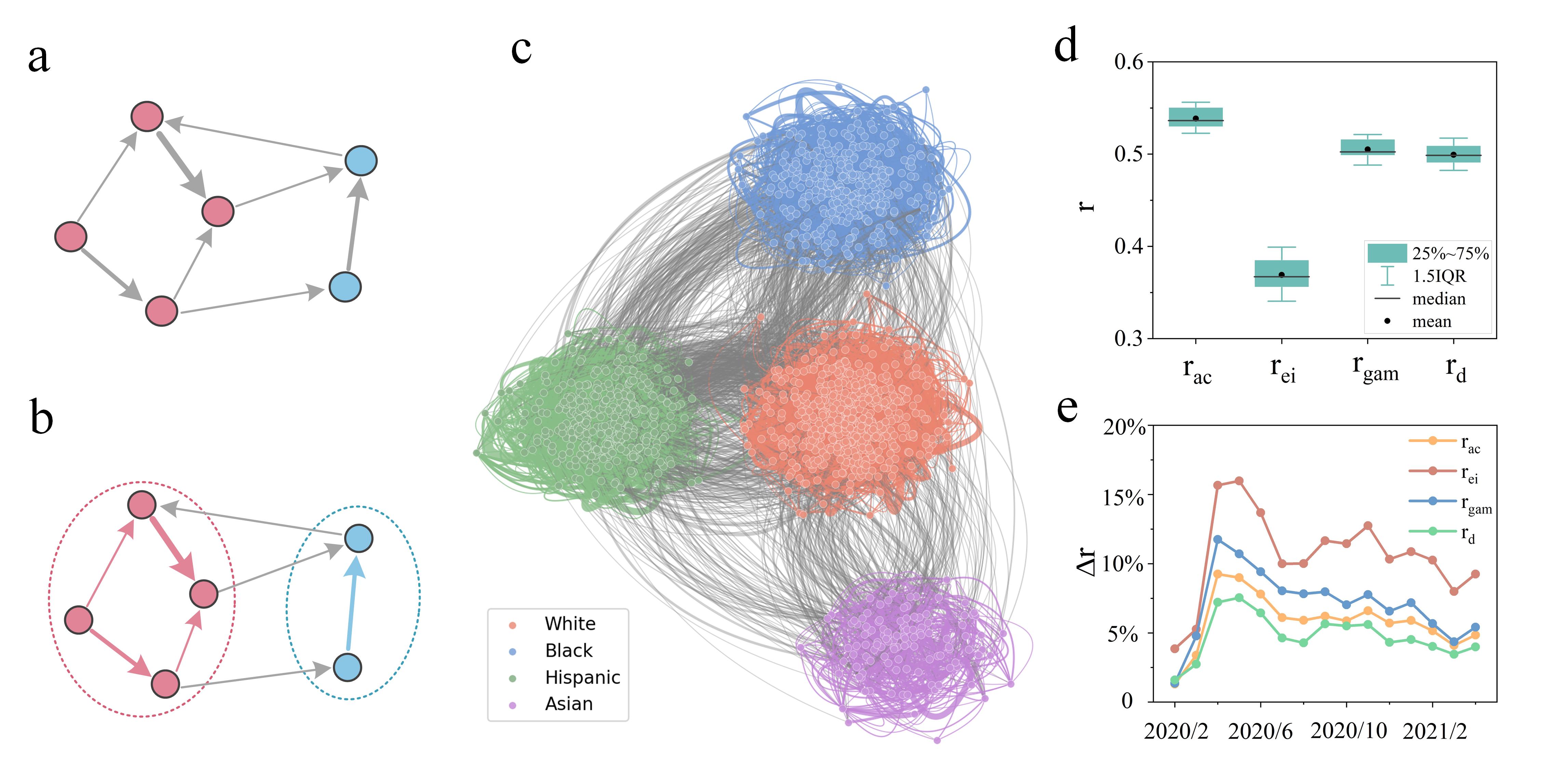}
    \caption{\textbf{a} A schematic diagram of racialized mobility network, nodes (Cts) in the network are colored by their majority racial group. Edge thickness scales with flow volume. \textbf{b} Network flows are partitioned into intra-group (colored edges) and inter-group (grey edges) components after aggregating tracts by racial majority. 
    \textbf{c} Visualisation of New York City's mobility network for February 2020, with nodes representing CTs grouped by majority racial composition. Edge thickness corresponds to monthly flow volumes. \textbf{d} Boxplots of four assortativity indices calculated monthly over the 14-month pre-pandemic period (January 2019-February 2020). All indices show consistently positive values (mean range: 0.369-0.538) with narrow interquartile ranges ($IQR < 0.022$), indicating stable baseline segregation patterns prior to COVID-19.
\textbf{e} Four assortativity indices ($r_{ac}$, $r_{ei} $, $r_{gam}$, $r_{d}$) showed significant increases during the pandemic compared to pre-pandemic baseline levels, following the implementation of containment measures.}
	\label{fig:fig2}
\end{figure}

In this study, we construct the racialized mobility network $G=(V,E)$ treats census tracts as nodes, with edges weighted by monthly visitor flows (Supplementary Note 3.1). Each node is annotated by its dominant racial group. Subsequently, we partition network flows into intra-group and inter-group components after aggregating tracts by racial majority, enabling comparative analysis of flow segregation patterns across racial categories: intra-group versus inter-group flows (see \hyperref[fig:fig2]{Fig.2a-b}). \hyperref[fig:fig2]{Fig.2c} visualizes the mobility network of New York City's mobility network for February 2020, where many thick edges appear in intra-group components connecting census tracts of the same racial group. When nodes are grouped into groups based on their categories, the network demonstrates statistically significant community structure (modularity $Q = 0.20$, $p < 0.001$ via $10,000$ random network permutations), indicating census tracts tending to form clusters based on racial composition\cite{newman2004finding}.

\subsection{Racial mobility mixing pattern}
\label{sec:mixing} 

We construct the mobility mixing matrix ($M_{m}$) to capture normalized flow patterns between racialized census tracts (see Methods section and Supplementary Note 3.2). The mobility mixing matrix for February 2020 (pre-pandemic) is shown in \hyperref[fig:matrices]{Table 1}, with other monthly results provided in \ref{apA}. Analysis of the pre-pandemic matrix reveals striking diagonal dominance across all racial categories. Intra-group flows constituted $69.27\%$ of total outflows. 
To systematically measure segregation intensity, we employ four assortativity indices adapted from network science: Assortativity Coefficient ($r_{ac}$), E-I index ($r_{ei}$), Gupta–Anderson–May index ($r_{gam}$), Diagonality index ($r_{d}$) (Supplementary Note 3.3) \cite{hilman2022socioeconomic, hu2019segregation}. These indices are widely employed to quantify intra-group flow concentration and evaluate preferential attachment within network communities. All four indices range from $-1$ (complete disassortativity) to 1 (perfect homophily), with positive values indicating intra-group flow preference. \hyperref[fig:fig2]{Fig.2d} shows boxplots of four assortativity indices calculated monthly over the 14-month pre-pandemic period (January 2019-February 2020). All indices show significant positive values (mean $\pm$ s.e.m.: $r_{ac} = 0.538 \pm 0.003$, $r_{ei} = 0.369 \pm 0.006$, $r_{gam} = 0.504 \pm 0.003$, $r_{d} = 0.498 \pm 0.003$), confirming pre-existing imbalance flows across racial groups. Such an imbalance implies that racialized mobility networks were governed by homophilic attraction, suggesting that movement patterns in New York City were systematically constrained by racial boundaries long before the pandemic.

During the pandemic, we track segregation dynamics by normalizing monthly index values to their corresponding pre-pandemic reference value (14-month mean from January 2019 to February 2020):
$\Delta r = \left({r_t}/{r_0} - 1\right) \times 100\%$, in which $r_t$ is the assortative index in $t$th month, $t_0$ is the pre-pandemic 14-month mean.
All assortativity indices exceeded their pre-pandemic baseline values after implementing the lockdown measures (March 2020), peaking in April-May 2020. $r_{ei}$ showed the most pronounced increase in intra-group flow retention ($16\%$ above pre-pandemic levels), potentially reflecting both heightened community solidarity and external exclusionary pressures. This surge in flow imbalance suggests that lockdown measures exacerbated pre-existing mobility disparities, restricting cross-racial movement. 

\begin{figure}[t]
\centering

\newcolumntype{C}{>{\centering\arraybackslash}p{1.2cm}}

\begin{subtable}{0.45\textwidth}
\centering
\caption{Table 1: Mixing Matrix ($M_m$) for February 2020} 
\begin{tabular}{@{} l | C C C C @{}}
\toprule
 & \textbf{White} & \textbf{Black} & \textbf{Hispanic} & \textbf{Asian} \\
\midrule
\textbf{White}  & 0.3285 & 0.0166 & 0.0255 & 0.0188 \\
\textbf{Black}   & 0.0610 & 0.1365 & 0.0301 & 0.0070 \\
\textbf{Hispanic}& 0.0750 & 0.0237 & 0.1910 & 0.0114 \\
\textbf{Asian}   & 0.0279 & 0.0034 & 0.0070 & 0.0367 \\
\bottomrule
\end{tabular}
\label{tab:table1}
\end{subtable}
\hfill
\begin{subtable}{0.45\textwidth}
\centering
\caption{Table 2: Row-Normalized Mixing Matrix ($M_r$) for February 2020}
\begin{tabular}{@{} l | C C C C @{}}
\toprule
 & \textbf{White} & \textbf{Black} & \textbf{Hispanic} & \textbf{Asian} \\
\midrule
\textbf{White}  & 0.8437 & 0.0425 & 0.0655 & 0.0482 \\
\textbf{Black}   & 0.2599 & 0.5820 & 0.1282 & 0.0299 \\
\textbf{Hispanic}& 0.2492 & 0.0787 & 0.6342 & 0.0378 \\
\textbf{Asian}   & 0.3720 & 0.0450 & 0.0937 & 0.4893 \\
\bottomrule
\end{tabular}
\label{tab:table2}
\end{subtable}

\vspace{0.5cm}

\begin{subtable}{0.45\textwidth}
\centering
\caption{Table 3: Baseline Segregation Matrix ($M_{bs}$) for February 2020}
\begin{tabular}{@{} l | C C C C @{}}
\toprule
 & \textbf{White} & \textbf{Black} & \textbf{Hispanic} & \textbf{Asian} \\
\midrule
\textbf{White}  &  0.4416 & -0.2143 & -0.1856 & -0.0416 \\
\textbf{Black}   & -0.1422 &  0.3251 & -0.1229 & -0.0599 \\
\textbf{Hispanic}& -0.1529 & -0.1782 &  0.3831 & -0.0520 \\
\textbf{Asian}   & -0.0301 & -0.2118 & -0.1574 &  0.3994 \\
\bottomrule
\end{tabular}
\label{tab:table3}
\end{subtable}
\hfill
\begin{subtable}{0.45\textwidth}
\centering
\caption{Table 4: Gravity-Based Segregation Matrix ($M_{gs}$) for February 2020}
\begin{tabular}{@{} l | C C C C @{}}
\toprule
 & \textbf{White} & \textbf{Black} & \textbf{Hispanic} & \textbf{Asian} \\
\midrule
\textbf{White}  &  0.2518 & -0.0938 & -0.0817 & -0.0211 \\
\textbf{Black}   & -0.0552 &  0.1607 & -0.0606 & -0.0337 \\
\textbf{Hispanic}& -0.0043 & -0.0434 &  0.0850 & -0.0201 \\
\textbf{Asian}   & -0.0167 & -0.1071 & -0.1135 &  0.2318 \\
\bottomrule
\end{tabular}
\label{tab:table4}
\end{subtable}

\label{fig:matrices}
\end{figure}


\subsection{Baseline Mobility Segregation}
 
To disentangle the confounding effects of demographic heterogeneity on mobility segregation measurement, we introduce a baseline null matrix $M_b$ to control for demographic distribution (see Methods section and Supplementary Note 4.1). We subtract $M_b$ from $M_r$ to yield the baseline segregation matrix $M_{bs}$, which serves as a residual matrix (see Methods section and Supplementary Note 5.1). \hyperref[fig:matrices]{Table 3} shows that the diagonal values (the self-segregation signal) in $M_{bs}$ greatly exceed 0, while the non-diagonal elements (the cross-group mixing signal) are almost all significantly below 0, indicating intra-racial mobility flows significantly exceeding the ideal mixing scenario and inter-racial flows systematically underperforming relative to the theoretical ideal.

By combining the self-segregation signal and cross-group mixing signal in the baseline segregation matrix, we propose two indices to distill these residuals into interpretable indices: the group-specific baseline index $S_b^{i}$ for racial group $i$, measuring its net isolation relative to demographic expectations; and the citywide baseline segregation index $S_b$, measuring average racial segregation across the urban mobility network (see Methods section). Two indices all range from -1 (excessive integration) to 1 (complete segregation) in practice. From January 2019 to February 2020 before the pandemic, the average value of $S_b^{i}$ is $0.537 \pm 0.003$, $0.516 \pm 0.004$, $0.526 \pm 0.003$, $0.443 \pm 0.003$ (mean $\pm$ s.e.m.) for White, Black, Hispanic, and Asian respectively, and $S_b$ is $0.504 \pm 0.003$ (see \hyperref[fig:fig3]{Fig.3a}). These values significantly exceed the ideal demographic parity baseline of zero, indicating substantial pre-existing mobility segregation across all racial groups prior to the pandemic. To quantify pandemic-induced changes, we compute percentage change $\Delta S_{b} = \left({S_{b}(t)}/{S_{b}(0)} - 1\right) \times 100\%$, ${S_{b}(t)}$ and ${S_{b}(0)}$  represent the $t$th month index value and the pre-pandemic 14-month mean. As shown in \hyperref[fig:fig3]{Fig.3b}, all segregation indices showed consistent increases during the pandemic. The peak amplification occurred in April 2020 (peak lockdown period), while the lockdown disproportionately affected minority groups: gravity-adjusted segregation surged by $14.2\%$ for Black and $11.3\%$ for Hispanic communities compared to pre-pandemic baselines. Our matrix comparison reveals that containment measures systematically exacerbated racial segregation by reinforcing intra-group mobility or suppressing cross-group movement. By comparing the matrixes of February 2020 and April 2020 (see \hyperref[tab:table3]{Table 3} and Table S4 in Supplementary Note),
intra-group interaction (diagonal elements) increased significantly for most groups except the White group, which slightly reduce from 0.4416 to 0.4036. Black group experienced the largest surges (from 0.3251 to 0.4104). The $S_b$ for White increased, which is primarily attributed to a significant reduction in mobility flow to predominantly White communities (off-diagonals in $M_{bs}$ declined, e.g., Black$\rightarrow$White value dropped from -0.0301 to -0.1099), revealing asymmetric reinforcement of racial mobility segregation.

\setcounter{figure}{2}

\begin{figure*}[t!]
    \centering 
    \includegraphics[width=13 cm,trim=11 0 0 10,clip]{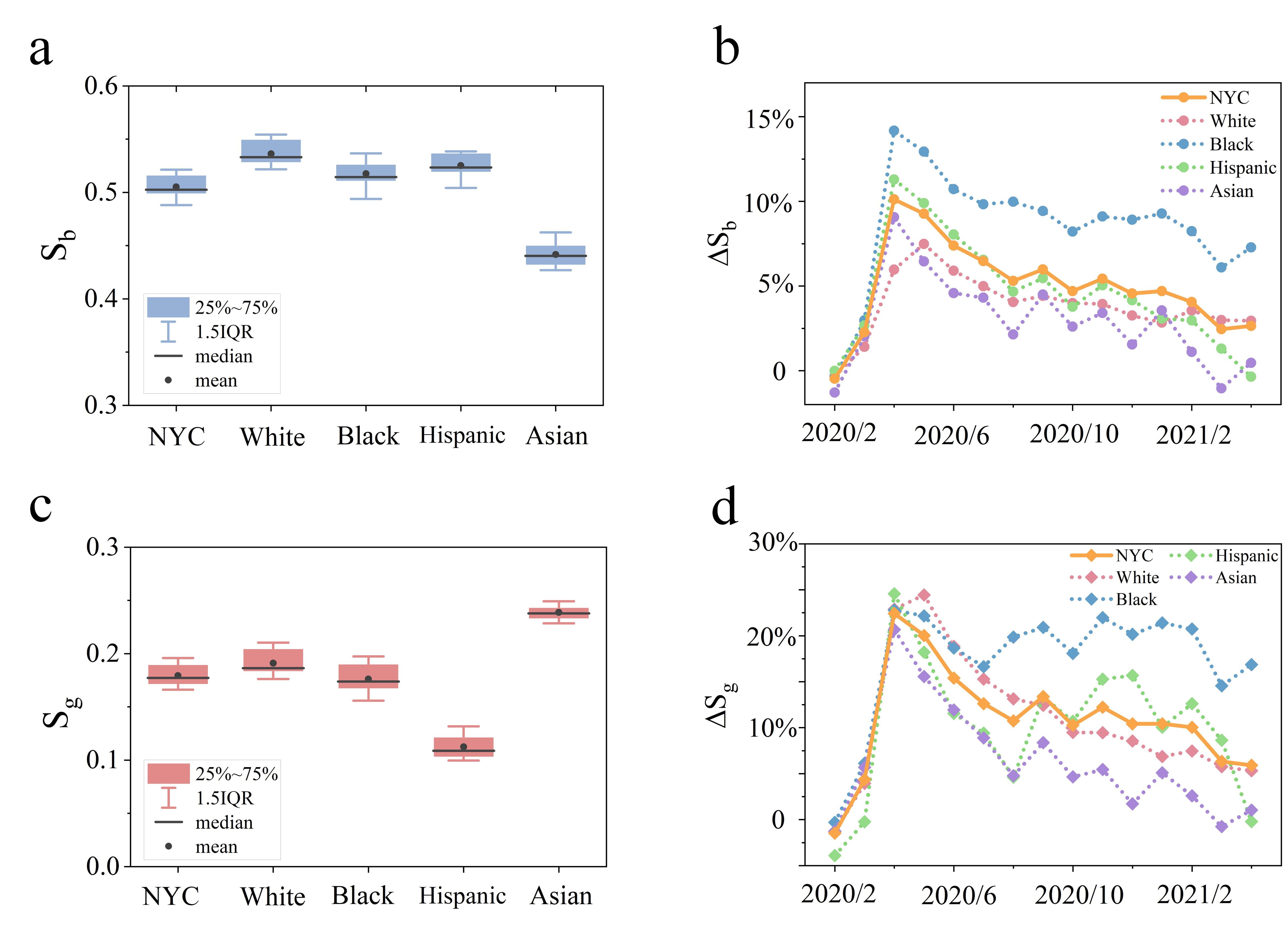}
    \caption{
\textbf{a} Boxplots of baseline segregation indices ($S_b^i$ and $S_b$) across racial groups during the 14-month pre-pandemic period, with all values significantly positive ($p<0.001$, Wilcoxon signed-rank test) with narrow interquartile ranges ($IQR \leq 0.017$), indicating persistent mobility segregation.
\textbf{b} Monthly fluctuations in $S_b$ and $S_b^i$ during the pandemic, showing increased segregation during lockdown periods. 
\textbf{c} Boxplots of gravity-based segregation indices ($S_g$ and $S_g^i$). The mean values exceed $0.113$, demonstrating reduced but still significant segregation ($p<0.001$), with narrow interquartile ranges ($IQR \leq 0.018$). \textbf{d} Comparative temporal patterns of gravity-adjusted segregation ($S_g$ and $S_g^i$), revealing similar pandemic amplification effects as baseline measures but with greater magnitude changes.}
	\label{fig:fig3}
\end{figure*}

Above analysis demonstrates that racial segregation in mobility patterns was a persistent feature of New York City's urban dynamics both before and during the pandemic, with the lockdown measures systematically exacerbating existing disparities across all groups. Critically, the baseline segregation matrix $M_{bs}$ quantifies the distance between observed mobility patterns and an idealized city where racial interactions would occur in exact proportion to neighborhood distributions (following the baseline null matrix)---a theoretical benchmark of demographic equity in urban mobility. The consistently positive $S_b$ values reveal that actual mobility networks systematically deviate from this equitable ideal, with intra-group flows exceeding and cross-group flows falling short of demographic expectations. This deviation metric essentially measures the ``segregation gap"—how far real-world mobility departs from a perfectly integrated urban system.

\subsection{Gravity-adjusted Mobility segregation}
\label{GRS}

While the baseline segregation index reveals systemic deviations from demographic parity, it conflates two distinct mechanisms: (1) passive exposure---increased co-racial encounters attributable solely to residential clustering patterns (reflecting structural segregation), and (2) active homophily---preferential flows toward co-racial destinations independent of spatial configuration (reflecting preferential segregation) \cite{mcpherson2001birds}. 
To disentangle active homophily from the confounding effects of passive exposure, we introduce a gravity-based null model that accounts for the natural decay of mobility flows with distance and construct a robust null expectation of mobility patterns (gravity-based null matrix $M_g$) driven purely by demographic and spatial constraints (see Methods section). Subtracting $M_g$ from the observed row-normalized mixing matrix $M_r$ yields the gravity-adjusted segregation matrix $M_{gs}$, whose elements reveal racial biases in mobility after accounting for geographic barriers. As shown in \hyperref[fig:matrices]{Table 4}, compared to the baseline segregation matrix $M_{bs}$, the values in $M_{gs}$ are systematically closer to zero—indicating that a significant portion of mobility segregation stems from the spatial clustering of racial groups rather than active homophily. However, the consistent pattern of positive diagonal elements and most negative off-diagonal elements in $M_{gs}$ confirms that active mobility preferences for same-race destinations persist beyond spatial explanations. 

Notably, while both matrices (\hyperref[fig:matrices]{Table 3} and \hyperref[fig:matrices]{Table 4}) show diagonal dominance, diagonal values reorder dramatically: from White (0.4416) $>$ Asian (0.3994) $>$ Hispanic (0.3831) $>$ Black (0.3251) in $M_{bs}$ to Asian Asian (0.2279) $>$ White (0.2054) $>$ Black (0.1024) $>$ Hispanic (0.0344) in $M_{gs}$. Their starkly divergent rankings reveal a nuanced segregation dynamic: the group appearing most self-segregated under demographic controls (White: $0.4416$) becomes only moderately segregated when spatial constraints are accounted for ($0.2054$), while the group with intermediate demographic segregation (Asian: $0.0.3994$) emerges as the most intensely self-segregated population ($0.2279$) after distance adjustment. Asian communities retain $57\%$ of their baseline self-segregation signal (0.3994$\rightarrow$0.2279) after spatial adjustment, suggesting their mobility patterns are actively sustained by cultural affinity and institutional completeness (e.g., linguistic networks, ethnic institutions) rather than passive residential clustering. Most crucially, the Hispanic experience presents a striking contrast---their segregation intensity plunges from 0.3831 (ranked 3rd in $M_{bs}$) to just 0.0334 (lowest in $M_{gs}$), a $87\%$ reduction that suggests Hispanic self-segregation derives primarily from residential clustering, with minimal additional active homophily.
This reversal exposes how conventional metrics conflate two fundamentally distinct phenomena---structural segregation embedded in neighborhood geography versus preferential segregation expressed through active homophily mobility behaviors. It demonstrates that relying solely on demographic factors may severely misestimated level of segregation, while gravity-adjusted segregation matrix helps uncover these hidden segregation patterns. 

Similar to the baseline segregation indices, we introduce gravity-adjusted segregation indices: the group-specific index $S_g^i$ and the citywide index $S_g$ (see Methods section). $S_g^i$ measures net segregation after controlling for both demographics and distance, with the average pre-pandemic values being $0.191 \pm 0.003$, $0.177 \pm 0.003$, $0.113 \pm 0.003$, $0.240 \pm 0.002$ (mean $\pm$ s.e.m.) for White, Black, Hispanic and Asian (see \hyperref[fig:fig3]{Fig.3c}). $S_g$ averages $0.180 \pm 0.003$, significantly lower than the baseline segregation indice ($S_b = 0.504$, $p<0.001$) yet still substantially above zero ($p<0.001$ via permutation tests). This confirms that while geographic proximity explains a portion of observed mobility segregation, a strong residual effect persists: individuals actively favour destinations with higher co-racial proportions and avoid cross-racial tracts beyond what spatial distance alone would predict.
This residual preference suggests active homophily mechanisms operate independently of urban spatial structure, indicating non-spatial mechanisms---like socio-cultural affinities (shared language/religion) and institutional contexts (service accessibility/policing patterns)---play a significant role in shaping segregated mobility patterns. The striking variation in segregation indices further corroborates these patterns: while Asian communities exhibit the lowest baseline segregation ($S_b^i = 0.443$), their gravity-adjusted index ($S_g^i = 0.240$) emerges as the highest—a reversal consistent with the $M_{bs}$ to $M_{gs}$ transition.

Subsequently, we calculate the percentage change of gravity-based segregation indices: $\Delta S_g^i$ and $ \Delta S_g$. During the pandemic, the gravity-adjusted segregation indices revealed a striking amplification, exhibiting both greater magnitude and persistence compared to baseline measures (see \hyperref[fig:fig3]{Fig.3d}). At the peak of lockdown restrictions in April 2020, the gravity-adjusted metrics showed an average increase of $22.72\%$ above pre-pandemic levels, substantially higher than the $10.39\%$ rise observed in baseline indices. Notably, Black communities exhibited the most persistent elevation in segregation levels throughout the pandemic, with gravity-adjusted indices maintaining an average increase of $20.86\%$ above pre-pandemic baselines—substantially higher than other racial groups and reflecting enduring mobility constraints within these neighborhoods. The systematic excess in gravity-adjusted fluctuations provides compelling evidence that pandemic measures disproportionately affected conscious travel decisions rather than just amplifying existing structural segregation embedded in residential clustering. These findings demonstrate that conventional demographic-based metrics may underestimate both the magnitude and duration of preferential segregation effects when populations face mobility constraints. Consequently, interventions targeting mobility-based segregation must take into account the gravity-adjusted metrics during crises to dynamically calibrate community outreach or resource allocation.  

\subsection{Exposure-Based mobility Segregation Dynamics}

\begin{figure}[t!]
    \centering 
    \includegraphics[width=16.5 cm,trim=0 0 0 0,clip]{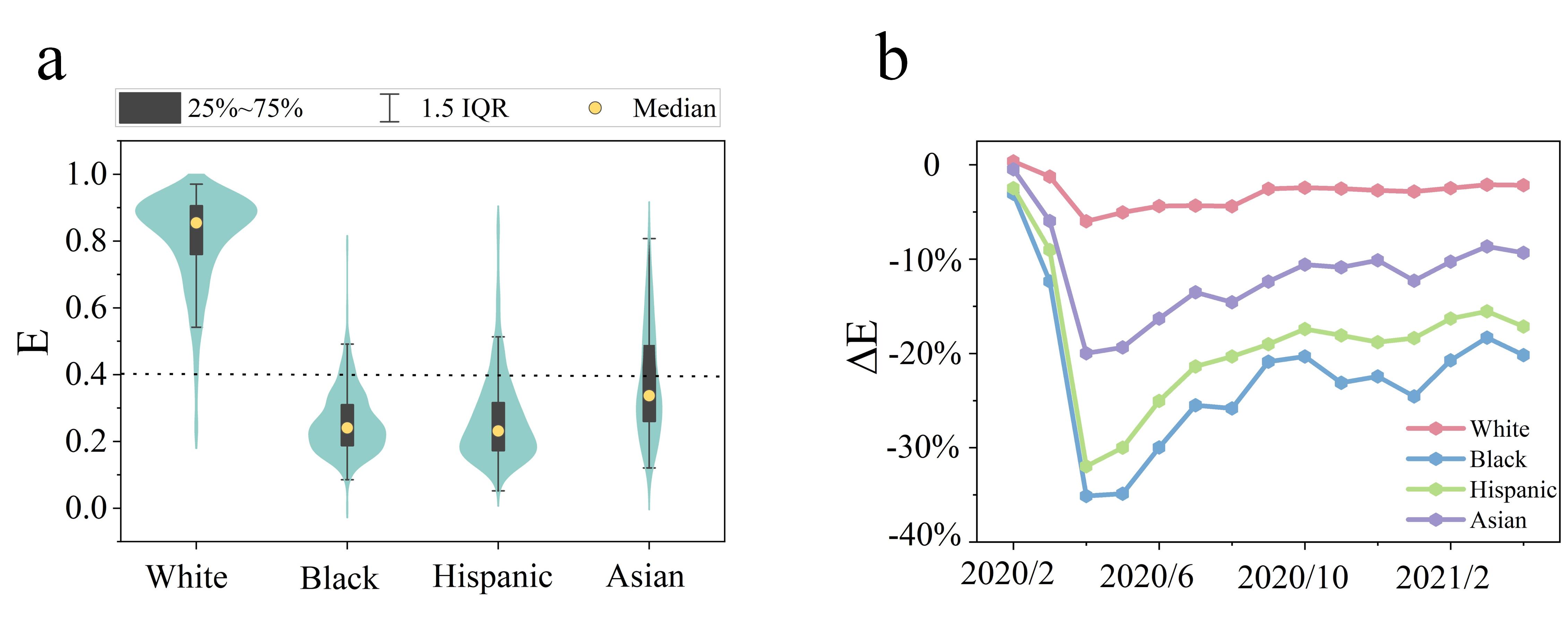}
    \caption{
\textbf{a} Violin plots of February 2020 exposure rates by racial group, revealing significantly higher mean exposure for White-majority CTs compared to Black, Hispanic, and Asian-majority CTs. The dotted line is the proportion of White-majority tracts to tracts dominated by all racial groups ($0.402$).
\textbf{b} Percent change in average exposure rates of each group relative to pre-pandemic (14-month mean before pandemic) across racial groups over time. White-majority CTs show the smallest decline (peaking at $-6\%$) and fastest recovery, while Black, Hispanic, and Asian-majority CTs sustained substantially larger reductions (exceeding more than $-20\%$) with slower rebound patterns.}
	\label{fig:fig4}
\end{figure}

To quantify the asymmetric nature of racial isolation in New York City, we analyze exposure probabilities to White-majority census tracts—a critical dimension of segregation in which minority groups face disproportionate barriers to accessing White-dominated spaces\cite{wang2018urban}. The raw exposure rate for the $n$th CT is defined as: $E_n = {F_{n \to W}}/{F_{n\rightarrow}}$ in which $F_{n \to W}$ is the outflows directed to White-majority tracts from the $n$th CT, $F_{n\rightarrow}$ is the total outflows of the $n$th CT. 
The proportion of White-majority tracts to tracts dominated by all racial groups is set as the benchmark ($0.402$), indicating the ideal exposure rate of each tract under a hypothetical equilibrium condition. As shown in \hyperref[fig:fig4]{Fig.4a}, the distribution of exposure rates reveals stark racial asymmetries before the pandemic. White-majority tracts exhibit near-universal access to their own neighborhoods (median exposure rate = 0.85), while minority-dominated tracts face severe exclusion from White spaces, with median exposure rates of 0.24 (Black), 0.23 (Hispanic), and 0.34 (Asian). It is notable that the median for the other three groups are all below the benchmark, while the White group is well above it. These pre-pandemic disparities demonstrate how structural advantages in mobility networks systematically privileged White-dominated spaces, creating unequal access to opportunity structures across racial groups even before COVID-19 disruptions. These results align with findings from Wang et al. \cite{wang2018urban}, who similarly documented structural advantages for White group exposing to White-dominated spaces in American cities.

These disparities intensified during the pandemic, \hyperref[fig:fig4]{Fig.4b} displays the percent change in average exposure rates relative to pre-pandemic (14-month mean before pandemic) across racial groups over time. In April 2021, the average exposure to White tracts declined by $35.1\%$ for Black residents, $31.2\%$ for Hispanics, and $20.0\%$ for Asians, compared to only $6.0\%$ for White residents. Crucially, White exposure rates rebounded to $97.5\%$ of baseline by September 2020, while Black and Hispanic groups remained $20-25\%$ below pre-pandemic levels—a divergence suggesting that crisis-period mobility restrictions disproportionately entrenched minority isolation from White social. This exposure-based analysis underscores how segregation operates directionally in urban networks: while all groups reduced cross-racial mobility during the pandemic, White communities retained significantly higher access to White-majority spaces. These findings underscore the need for equity-focused mobility policies during public health crises. Due to the existing structure of residential isolation, blanket restrictions may inadvertently exacerbate existing racial disparities in spatial access and social connectivity, for example: the access routes for black people to medical resources may be cut off by containment measures. Effective interventions must therefore implement racial impact assessments before enacting mobility policies, coupled with compensatory connectivity measures (e.g., targeted shuttle services linking minority enclaves to vaccine sites or job hubs).

\subsection{Homophily Gravity Model}
\label{model_result}

\begin{figure}[t!]
    \centering 
    \includegraphics[width=15cm,trim=0 0 0 5,clip]{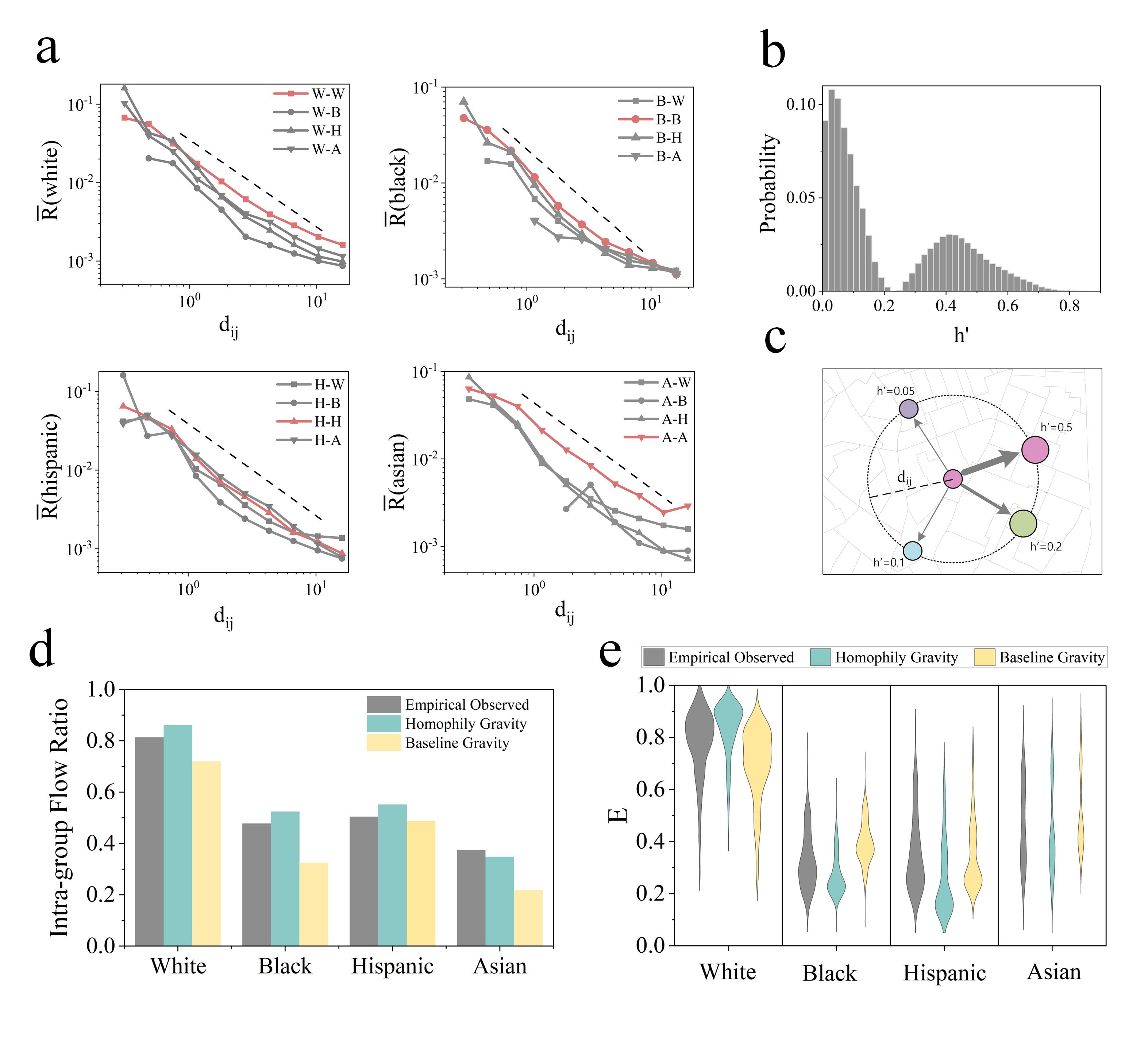}
    \caption{\textbf{a} Average flow proportion $\bar R$ from CT $i$ in a racial group (White: W; Black: B; Hispanic: H, Asian: A) traveling to CT $j$ in another group with the distance $d_{ij} $. $\bar R$ decays with $d_{ij}$ with a slope $\bar R \propto d_{ij}^{\eta}$, in which $\eta \in[-1.2,-1.1]$. The curve representing flows between the same racial groups predominantly resided above the curves for flows between different groups across most distance intervals. \textbf{b} The distribution of the primary racial similarity score $h'$ exhibits a bimodal pattern. This 0.25 inflection point thus serves as a natural separator: values below 0.25 predominantly represent tract pairs with divergent dominant racial compositions, while those to its right demonstrate that dominant race is the same. \textbf{c} Homophily Gravity model (HGM) illustration: there is more flow between the two CTs with a larger primary racial similarity score $h'$, when controlling for population size and centroid-to-centroid distance. Different colors represent different dominant races, and the size of the points and edges indicates the flow volume. \textbf{d} Comparison of intra-group flow ratios: empirical vs. model predictions. HGM better approximates empirical values for White, Black, and Asian communities, but shows limited improvement for Hispanic communities. \textbf{e} Model fidelity in replicating racial exposure disparities. Violin plots show distributions of exposure rates to White-majority tracts. HGM achieves consistently lower Wasserstein distances (WD) values (indicating better fit) for White (WD$=0.058$ vs. $0.071$ in BGM), Black (WD$=0.056$ vs. $0.080$ in BGM), and Asian (WD=$0.038$ vs. $0.057$ in BGM) tracts, except Hispanic tracts (WD$=0.065$ vs. $0.022$ in BGM).}
	\label{fig:fig5}
\end{figure}

Prior results established a clear pattern: flows between CTs dominated by the same racial group generally exceeded flows between CTs dominated by different groups. To investigate how this preference operates across geographical space, we analyzed the distance decay profiles of proportional flows. Specifically, we examined the average proportion of outflows $\bar R$ from CTs dominated by each racial group (White, Black, Hispanic, Asian) directed towards CTs dominated by each of the four groups, as a function of the Euclidean distance $d_{ij}$ (in kilometers) between CT centroids. \hyperref[fig:fig5]{Fig.5} shows the empirical results for February 2020. As anticipated by fundamental spatial interaction principles\cite{tobler1970computer}, all flow proportion curves exhibited a distinct power-law decay pattern with increasing distance $\bar R \propto d_{ij}^{\eta}$, in which $\eta \in [-1.1, -1.2]$). Crucially, the curve representing flows to destinations dominated by the same racial group as the origin predominantly resided above the curves for flows to destinations dominated by the other three groups across most distance intervals, though this dominance was less pronounced at the head of the curve (shorter distances). This pattern manifested most prominently among White and Asian groups, while the two groups also exhibited the strongest diagonal elements in the gravity-adjusted segregation matrix. In contrast, the advantage for same-race flow curve was markedly weaker among Hispanic communities—consistent with their isolation stemming predominantly from residential clustering rather than active mobility preferences. These findings suggest that while distance universally reduces mobility between tracts, same-race flows exceed cross-race flows at most spatial scales, pointing to an additional, race-driven factor influencing urban movement patterns.

Spatial interaction patterns are known to be influenced not only by geographic distance and population sizes, but also by socio-demographic affinities, such as shared race, ethnicity, religion, or socioeconomic status\cite{hu2019segregation,moro2021mobility,yabe2025behaviour}---a phenomenon broadly termed homophily. While classical gravity models effectively capture the decay of flows with distance and the scaling with origin/destination masses \cite{tobler1970computer, barthelemy2011spatial}, they typically lack explicit parameters to quantify how such social similarities or dissimilarities systematically modulate mobility preferences. This limitation hinders a comprehensive understanding of segregation dynamics embedded within urban movement networks, particularly across multiple potential dimensions of social identity. The empirical evidence in our study necessitates an expansion of these models to incorporate social affinities as a fundamental dimension of mobility friction. This is particularly critical in cities characterized by entrenched socio-cultural and economic disparities, where multidimensional segregation creates persistent structural barriers to intergroup connectivity—operating synergistically with geographical distance to fragment urban mobility networks. To bridge this methodological gap, we propose the Homophily Gravity Model (HGM)---a novel framework formally incorporating social affinity as a fundamental dimension of mobility friction (see Methods section and Supplementary Note 6.1). Within this generalized architecture, the present study operationalizes the HGM specifically for racial homophily. We first introduce a racial similarity parameter $h$, which quantifies dominant race alignment between CTs. The parameter construction proceeds through two conceptual stages: First, calculating a primary racial similarity score ($h'_{ij}$) based on dominant racial group proportions (Supplementary Note 6.2), to ensure maximal values when tracts share both dominant racial groups and their high proportions. \hyperref[fig:fig5]{ Fig.5b} illustrates the probability distribution of the primary racial similarity score $h'$, which exhibits a bimodal pattern, with a trough at $0.25$ that effectively demarcates distinct regimes of inter-tract relationships. \hyperref[fig:fig5]{Fig.5c} illustrates the Homophily Gravity Model (HGM) mechanism, where flow volume between census tracts (CTs) scales with their primary racial similarity score $h'_{ij}$, after controlling for population size and distance. Second, applying an exponential transformation $h_{ij} = e^{h'_{ij}}$ to enhance discrimination between integrated and segregated CTs. To quantify racial homophily in mobility flows, we integrate $h_{ij}$ as a multiplicative factor into the classic gravity model (see Methods section) \cite{barthelemy2011spatial, nie2023examining}. The estimated exponent $\gamma$ measures how strongly racial homophily modulates flows. 

\setcounter{table}{4}

\begin{table}[t!]
  \begin{minipage}{0.4\textwidth}
    \centering
    \caption{Gravity Model Specifications (November 2019 - February 2020)}
    \label{tab:model_comparison_left}
    \begin{tabular}{lcc}
  \toprule
      \textbf{Parameter} & \textbf{Baseline (w/o $h_{ij}$)} & \textbf{Full (w/ $h_{ij}$)} \\
  \midrule
  \multicolumn{3}{l}{\textbf{Conventional Parameters}} \\
  $\lambda$ & 5.304***  & 2.316***  \\
            & (0.028)   & (0.012)   \\
  $\alpha$  & 2.303***  & 2.094***  \\
            & (0.003)   & (0.003)   \\
  $\beta$   & 1.174***  & 1.232***  \\
            & (0.001)   & (0.001)   \\
  $\delta$  & 1.230***  & 1.110***  \\
            & (0.001)   & (0.001)   \\
  \midrule
  \multicolumn{3}{l}{\textbf{Racial Parameter}} \\
  $\gamma$  & \multicolumn{1}{c}{--} & 2.259*** \\
            &                       & (0.007)  \\
  \midrule
  Observations & 937,705 & 937,705 \\
  $R^2$       & 0.451    & 0.511     \\
  Adjusted $R^2$ & 0.451  & 0.511     \\
\bottomrule
\multicolumn{3}{l}{\footnotesize \textit{Notes: *$p < 0.1$; **$p < 0.01$; ***$p < 0.001$.}} \\
\end{tabular}
  \end{minipage}
  \hspace{2cm}
  \begin{minipage}{0.4\textwidth}
    \centering
    \caption{Gravity Model Specifications (March 2020 - June 2020)}
    \label{tab:model_comparison_right}
\begin{tabular}{lcc}
  \toprule
      \textbf{Parameter} & \textbf{Baseline (w/o $h_{ij}$)} & \textbf{Full (w/ $h_{ij}$)} \\
  \midrule
  \multicolumn{3}{l}{\textbf{Conventional Parameters}} \\
  $\lambda$ & 43.913*** & 8.816*** \\
            & (0.194)   & (0.058)   \\
  $\alpha$  & 1.468***  & 1.632***  \\
            & (0.004)   & (0.003)   \\
  $\beta$   & 1.187***  & 1.302***  \\
            & (0.002)   & (0.002)   \\
  $\delta$  & 1.213***  & 1.019***  \\
            & (0.002)   & (0.002)   \\
  \midrule
  \multicolumn{3}{l}{\textbf{Racial Parameter}} \\ 
  $\gamma$  & \multicolumn{1}{c}{--} & 2.903*** \\
            &                       & (0.010)  \\
  \midrule
  Observations & {727,269} & {727,269} \\
  $R^2$       & 0.332    & 0.395     \\
  Adjusted $R^2$ & 0.332  & 0.395     \\
\bottomrule
\multicolumn{3}{l}{\footnotesize \textit{Notes: *$p < 0.1$; **$p < 0.01$; ***$p < 0.001$.}} \\
\end{tabular}
    
\end{minipage}
\vspace{-0.1cm}
\end{table}

To ensure robust quantification of race-modulated gravitational forces, we calibrated both Homophily Gravity Model (HGM) and Baseline Gravity Model (BGM) using trips between census tracts with centroid distances $d_{ij} > 1$ km (Supplementary Note 6.3). This threshold preserves $76\%$ of inter-tract flows while excluding short-distance regimes where racial homophily signals are confounded by proximity constraints.
\hyperref[tab:model_comparison_left]{Table 5} illustrates the results when we calibrated the Homophily Gravity Model and the Baseline Gravity Model (BGM) using pre-pandemic data (November 2019-February 2020). In the HGM specification, traditional flow parameters showed expected patterns: origin outflow ($\alpha = 2.30$, SE $= 0.003$) had greater influence than destination inflow ($\beta = 1.17$, SE $= 0.001$), while distance decay ($\delta = 1.22$, SE $= 0.001$) confirmed spatial friction. Notably, the inclusion of racial similarity parameter ($h_{ij}$) significantly enhanced the gravity model's explanatory power, as evidenced by a $0.06$ (from $0.451$ to $0.511$) improvement in adjusted $R^2$ compared to the BGM specification (without $h_{ij}$), alongside a statistically significant exponent $\gamma$ of $2.25$ ($p<0.001$). This enhancement confirms that racial alignment captures unique variance in mobility patterns beyond traditional geographic and demographic factors. \hyperref[fig:fig5]{Fig.5d} compares intra-group flow ratios derived from empirical observations, the HGM, and the BGM across racial groups. While the BGM systematically underestimates intra-group ratio (by $11.5-41.5\%$), the HGM demonstrates significant improvement for most groups. We quantify model improvement using the relative improvement rate, which measures the percentage reduction in prediction error compared to the baseline model: $(1 - |\hat{I}_{\text{HGM}} - I|/|\hat{I}_{\text{BGM}} - I|) \times 100\%$
where $\hat{I}$ represents model's prediction, and $I$ is the observed value. The HGM achieves relative improvement rates of $49.9-83.0\%$ for White, Black, and Asian populations. The exception occurs for Hispanic communities where the baseline model's proximity to empirical values. Its weaker performance for Hispanic communities aligns with the distinct segregation mechanisms observed across racial groups. Our Analysis in Section \ref{GRS} indicates Hispanic isolation stems primarily from passive residential clustering rather than active homophily—a pattern our unified racial similarity parameter $h$ captures less effectively because it assumes consistent behavioral homophily across all groups. To quantify how well each model reproduces empirical exposure patterns, we compare their Wasserstein distances (WD) to observed data—--a robust metric measuring the minimum `work' required to transform one distribution into another \cite{vaserstein1969markov}. Our HGM systematically outperforms the BGM across White, Black, Asian groups (White: $0.071 \rightarrow 0.058$, Black: $0.080 \rightarrow 0.056$, Asian: $0.057 \rightarrow 0.038$). The sole exception also occurs for Hispanic tracts ($0.022 \rightarrow 0.065$) where the BGM's result is closer to empirical data than our model, consistent with their unique structural segregation dynamics discussed earlier. This metric confirms our model's superior capacity to capture directional isolation patterns, particularly for groups exhibiting active homophily (White/Black/Asian).

As shown in \hyperref[tab:model_comparison_right]{Table 6} (the modeling results for April-July 2020), incorporating racial homophily still significantly improved the model's explanatory power during pandemic, with $R^2$ increasing by $0.063$, $\gamma = 2.903$, $SE = 0.010$, $p < 0.001$. 
Notably, although $\gamma$ was higher during November 2019–February 2020 than in the pre-pandemic period, it did not remain elevated as expected. In July–October 2020, $\gamma$ dropped to 2.01—lower than the pre-pandemic value of 2.25—even though the model still achieved improved predictive accuracy ($\Delta R^2=+0.046$; see Table S11 in the Supplementary Note).
This contradiction - where macroscopic segregation indices showed increased racial homogeneity during the pandemic but $\gamma$ values decreased - can be explained by the complex spatial restructuring of mobility networks during the pandemic. While overall intra-racial flows increased at the aggregate level, New York City's localized containment measures created uneven effects across neighborhoods, with high-risk zones experiencing greater flow reductions between certain sub-communities within racial groups. This micro-scale heterogeneity meant that while some community pairs showed strengthened homophily effects, others exhibited weakened connections, leading to the observed fluctuation in the overall $\gamma$ coefficient. Crucially, the continued improvement in $R^2$ demonstrates that the $h_{ij}$ parameter indeed captures additional variance beyond what is explained by traditional geographic/demographic factors alone, even though $\gamma$ did not fully reflect the macro-level increase in racial segregation.

\section{Discussion and Conclusion}
\label{sec:discussion}

Understanding how systemic shocks reconfigure racial segregation in urban mobility networks is critical for equitable crisis response, as cities face compounding pressures from public health emergencies \cite{jia2020population}, climate disruptions \cite{masson2020urban,montfort2025systematic}, and technological transformations in transportation \cite{maheshwari2021will}. While prior research has extensively quantified segregation through point-to-point
visitation probabilities or co-presence metrics at specific locations \cite{hilman2022socioeconomic,moro2021mobility,wang2018urban}, the segregation embedded in citywide racialized flow networks and their dynamic changes in response to the impact of the pandemic remain insufficiently conceptualized. Moreover, prior work has often framed mobility segregation as a mere counterpoint to residential segregation—a binary opposition that oversimplifies its multidimensional nature. This approach neglects the complex interplay of structural constraints and agent-driven behaviors that collectively shape segregated movement patterns. This gap is consequential: mobility segregation operates not merely through neighborhood demographics but through latent subjective preferences in travel choices, with cascading effects on access to jobs, healthcare, and social capital. Leveraging large-scale mobility data from New York City, we deconstruct mobility segregation within racialized mobility networks from multi-dimensional perspective and interrogate how COVID-19 containment measures acted as a natural experiment to expose and amplify these hidden fissures.

Our research provides critical theoretical and empirical advances in analyzing racialized mobility networks under crisis conditions through three principal contributions. First, by constructing interracial mobility networks and developing a multidimensional analytical framework (encompassing mixing matrices, baseline segregation indices, gravity-adjusted metrics, and exposure rates), we uncover deeply entrenched segregation patterns in pre-pandemic urban flows. Crucially, our decomposition of mobility segregation into structural and preferential dimensions reveals that racial divisions transcend residential clustering---Asian communities, for instance, maintain persistent mobility homophily even after spatial adjustments, demonstrating how cultural affinity actively sustains segregated mobility patterns independently of geographic constraints. Second, longitudinal analysis of pandemic mobility data demonstrates that containment measures asymmetrically intensified segregation across all dimensions, with disproportionate effects: while Black and Hispanic communities experienced amplified structural isolation, White neighborhoods showed the most pronounced increase in active homophily, exacerbating pre-existing inequities in access to opportunity structures. Third, building on observed distance-decay patterns in racial flow proportions, we develop a Homophily Gravity Model (HGM) that achieves superior predictive accuracy, and replicate both intra-group mobility preferences and exposure rate to White-dominated spaces with greater fidelity. This modeling breakthrough formally establishes racial homophily as a trans-spatial force that systematically reorganizes urban mobility networks beyond the constraints of physical distance.

These findings carry implications for both research and policy. Methodologically, our dual-metric framework advances segregation analysis by disentangling structural constraints from active homophily in mobility behaviors, addressing a key limitation of traditional indices. Yet limitations remain: our data cannot distinguish between avoidance behaviors (e.g., fear of interracial contact) and solidarity-driven intra-group reliance, nor capture granular socioeconomic factors shaping mobility \cite{li2023assessing}.  
Our Homophily Gravity Model with the racial similarity parameter represents a pioneering attempt to formalize how racial preferences modulate mobility networks, bridging the gap between macroscopic segregation indices and micro-level mobility decisions. The modeling results demonstrate that traditional geographic/demographic factors alone cannot fully explain racialized movement patterns, and explicit modeling of racial homophily is essential for accurate mobility prediction in cities with severe racial segregation, which offers a refined lens for segregation dynamics in metropolitan areas.
However, exceptional fluctuations in racial homophily coefficients ($\gamma$) and modeling results of Hispanic communities reveal a limitation of our current framework: the assumption of uniform homophily effects across all sub-communities may oversimplify context-dependent variations in racial preferences. Future research could attempt to implement distinct $h$ coefficients for each racial group to capture how cultural affinity, historical segregation, and structural privilege differentially shape racial homophily in mobility patterns. In addition, the generalized architecture of the HGM provides a ready pathway for investigating how social affinity (e.g., racial, religious, or socioeconomic similarity) modulates spatial interactions. Realizing this potential requires deliberate design of $s_{ij}$ parameters that capture domain-specific friction mechanisms, whether in educational attainment, occupational segregation, or cultural alignment. These open questions highlight the need for future research extending the HGM architecture to investigate multi-dimensional segregation (e.g., defining $s_{ij}$ for income similarity, religious alignment, or linguistic proximity), enabling comparative analysis of how different social fractures shape urban connectivity.

Practically, policies aimed at fostering equitable urban resilience—such as transit investments linking segregated neighborhoods or mixed-use zoning---not only need to account for static residence segregation and mobility segregation mechanisms, but also structural segregation and preferential segregation. Our results also caution against one-size-fits-all containment measures, which risk cementing fragmentation. For policymakers, the gravity-adjusted segregation matrix ($M_{gs}$) and indices ($S_g$) offer diagnostic tools to prioritize interventions: communities with high baseline but low gravity-adjusted segregation (e.g., Hispanic tracts) require spatial equity measures (e.g., improved transit links to opportunity hubs), while those with persistent gravity-adjusted isolation (e.g., Asian tracts) may need cultural or institutional outreach to reduce behavioral barriers. Future research should extend this framework to other cities, segregation types (e.g., economy, religion) and shocks (e.g., earthquake, climate disasters) \cite{lu2012predictability, hyra2013metropolitan}, while integrating individual-level socioeconomic data to unpack the microfoundations of segregation. The pandemic’s legacy may include enduring shifts in how race structures mobility networks, with cascading effects on social cohesion and economic opportunity \cite{kaye2021economic}. This necessitates segregation-sensitive containment policies that proactively assess racial equity impacts before implementation---for example, by employing gravity-adjusted segregation metrics to predict how mobility restrictions might asymmetrically constrain minority access to essential services. Without such safeguards, well-intentioned public health measures risk cementing crisis-born segregation patterns into permanent urban infrastructures of exclusion. As cities face growing threats from climate change, political instability, and technological disruption, understanding mobility segregation as a dynamic network---rather than a static spatial pattern---will be essential for fostering resilient, inclusive communities \cite{masson2020urban,dia2020technology}. Our approach provides a foundation for this work, bridging the gap between segregation theory and the lived experience of division in daily movement.

\section{Methods}
\subsection{Data Description}

We utilize anonymized human mobility data in New York City covering the period from January 2019 to April 2021, which includes both pre-pandemic and pandemic phases, sourced from mobile device location records provided by SafeGraph and processed by Kang et al.\cite{kang2020multiscale}. The dataset aggregates daily visitor flows between census tracts. For our analysis, we aggregated these daily visitor flows into monthly totals, representing the cumulative number of visitors traveling between each census tract pair (a$\rightarrow$b) per month (see Supplementary Note 1.1 for more details). In the pre-pandemic period, New York City averaged approximately 24.4 million monthly inter-tract visitor flows, establishing a baseline for our subsequent analysis of mobility segregation. Following pandemic onset, monthly visitor flows plummeted to a trough of just 11.3 million in April 2020 - a nearly $53\%$ reduction from baseline levels.

To contextualize mobility patterns with demographic attributes, we integrate racial composition data from the 2020 Decennial Census (U.S. Census Bureau, \url{https://www.census.gov/}). These data provide tract-level population statistics, including: proportions of White, Black/African American, Asian, Hispanic/Latino, and other racial groups within each census tract. Census tract boundaries are harmonized with the mobility dataset to ensure consistent spatial units for analysis. While the mobility data span 2019–2021, the 2020 Census serves as the demographic anchor, as annual estimates for 2019–2021 exhibit negligible tract-level racial composition shifts \cite{us2020understanding}. Our analysis focuses on New York City's 2,327 census tracts (CTs). We selected CTs with a population size exceeding 100 and where a single racial group constitutes $\geq 50\%$ of the population. The $\geq50\% $threshold ensures that the dominant racial/ethnic group represents a socially salient community identity in segregation dynamics. This yielded four distinct categories: majority-White (631 CTs), majority-Black (403 CTs), majority-Hispanic (394 CTs), and majority-Asian (141 CTs), totaling 1,569 CTs ($67.4\%$ of all CTs in NYC). Details of tract selection can be found in Supplementary Note 1.3

\subsection{Mobility Mixing Matrix}

The mobility mixing matrix $M_{m}$ represents normalized flow probabilities from census tracts dominated by one racial group to tracts dominated by the same or different groups \cite{newman2003mixing}. Specifically, we aggregate all outflows from nodes (tracts) in racialized mobility network dominated by racial group $i$ to nodes dominated by group $j$, then normalize by the total outflows from all nodes (TotalOutflows). Formally, each element $m_{ij} \in M_{m}$ denotes: $m_{ij} = \text{Flows}(i \to j) / \text{TotalOutflows}$, where $i,j \in \{\text{White, Black, Hispanic, Asian}\}$ denote the four racial groups under study. To quantify the structural mixing patterns encoded in the mobility mixing matrix, we introduce several assortativity indices to quantify whether and to what extent links tend to connect nodes of the same type\cite{newman2003mixing,hu2019segregation}: Assortativity Coefficient ($r_{ac}$), E-I Index ($r_{ei}$)
, Gupta-Anderson-May Index ($r_{gam}$), Diagonality Index ($r_{di}$) (Supplementary Note 3.3). All four indices range between $-1$ and $1$, while $1$ indicates fully segregated mobility patterns and $-1$ represents complete disassortative mixing.

The values in mixing matrix are inherently skewed by demographic disparities, conflating population-size effects with true segregation patterns. To ensure equitable analysis of segregation patterns across racial groups, we construct a row-normalized mixing matrix $M_{r}$ where each element $m^r_{ij}$ represents the proportion of outflows from racial group $i$ that terminate in group $j$, relative to the total outflows of group $i$ ($\text{TotalOutflows}(i)$). $m^r_{ij}$ is given by: $m^r_{ij} = \text{Flows}(i \to j) / \text{TotalOutflows}(i)$.
By scaling each group's outflows to unity, our methodology deliberately neutralizes demographic asymmetries through flow normalization (Supplementary Note 3.4). 

\subsection{Null Mixing Matrix}
\label{null_model}

To distinguish mobility segregation patterns from confounding factors (including population-size disparities and residential clustering effects), we construct two null mixing matrices that serve as theoretical baselines:

\paragraph{Baseline Null Matrix}
The baseline null matrix $M_{b}$ is a special row-normalized mixing matrix assuming that racial mixing probabilities are strictly proportional to the geographic distribution of racial groups. Each element $m^b_{ij}$ in row $i$ ($i$ is an origin racial group) is identical and defined as:

\begin{equation}
     m^b_{ij} = \frac{N_j}{\sum\limits_{k=1}^4 N_k}
\end{equation}
where $N_j$ is the count of tracts dominated by group $j$ (Supplementary Note 4.1). 
The matrix therefore serves as a critical reference point that captures expected mobility patterns under conditions of perfect demographic neutrality, where no racial group exhibits preferential attachment or avoidance behaviors, and where spatial arrangements impose no additional constraints on mobility beyond the basic distribution of racial groups across the urban landscape. For example, if Hispanic-majority tracts comprise $25\%$ of all tracts, we assume that they would receive $25\%$ of flows from every origin group. 

\paragraph{Gravity-Based Null Matrix}

Taking into account the spatial limitations for urban residents' travel, we introduce a gravity-based null model and construct the gravity-based null matrix $M_{g}$  to incorporates spatial constraints to theoretical flows. The equation of the null gravity model takes the following form:

\begin{equation}
    F_{ij} = \lambda\frac{f_{i,out}^\alpha f_{j,in}^\beta}{d_{ij}^\delta }
\end{equation}

where $f_{i,out}$ and $f_{j,in}$ is outflow of origin tract $i$ and inflow of destination tract $j$ respectively; $d_{ij}$ denotes the Euclidean distance between the centroids of tracts $i$ and $j$, and $\lambda$, $\alpha$, $\beta$, $\delta$ are parameters fitted via maximum likelihood estimation. While traditional gravity models often use residential population as a proxy for location attractiveness, we employ the dynamic active population - the count of unique visitors to or from a location - which better captures the actual human activity patterns and temporal variations in destination attractiveness \cite{barthelemy2011spatial, nie2023examining}. Using the fitted gravity model, we generate a synthetic flow network where each link $F_{ij}$ represents the expected flow under spatial and demographic constraints (Supplementary Note 4.2). Subsequently, the gravity-based null matrix $M_{g}$ can be constructed from the synthetic network. This null model accounts for both demographic distributions and the friction of distance, providing a more realistic counterfactual than the proportional baseline. 

\subsection{Segregation Index}

To quantify racial segregation in mobility patterns, we propose two classes of indices derived by comparing observed mixing matrices against null models:

\paragraph{Baseline Segregation Index}
To evaluate segregation relative to demographic parity, we construct the baseline segregation matrix ($M_{bs}$) by subtracting the baseline null matrix ($M_b$) from the row-normalized mixing matrix ($M_r$): $M_{bs} = M_{r}-M_{b}$, which quantifies how observed mixing deviates from a society where mobility probabilities strictly follow tract composition. The elements of the baseline segregation matrix $M_{bs}$ reveal meaningful patterns about racial mobility preferences, while positive values $m^{bs}_{ij}>0$ indicate that the observed flows from group $i$ to group $j$ exceed expectations under demographic parity and negative values $m^{bs}_{ij}<0$ signify suppressed mobility below demographic expectations. In this study, the diagonal entries $m^{bs}_{ii}$ are defined as the self-segregation signal, measuring the intensity of within-group segregation. The off-diagonal elements $m^{bs}_{ij}$ ($i\neq j$) are defined as the cross-group mixing signal, capturing the mixing patterns between different racial groups. By combining two signals, we propose group-specific segregation indices for race $i$:

\begin{equation}
    S_b^{i} = \underbrace{m^{bs}_{ii}}_{\text{self-segregation}} 
- \underbrace{\frac{1}{3}\sum_{i \neq j} m^{bs}_{ij}}_{\text{cross-group mixing}}
\end{equation}
in which the self-segregation item quantifies group $i$'s self-segregation degree (avoidance of out-group mobility), and the cross-group mixing term captures the other groups' visitation preferences toward group $i$. This index measures the net segregation for group $i$ by contrasting its diagonal excess with off-diagonal deficits. Thus, the segregation index captures the combined effect of group $i$'s tendency to self-isolate and the aggregate reluctance of other groups to visit group $i$. This dual-term design operationalizes Schelling's theory of segregation dynamics\cite{1971Dynamic}. The citywide baseline segregation index is defined as the average of all group-specific indices:

\begin{equation}
   S_b = \frac{1}{4}\sum_{i=1}^4 m^{bs}_{ii} 
- \frac{1}{12}\sum_{i \neq j} m^{bs}_{ij} = \frac{1}{4}\sum_{i=1}^4 S_b^{i}
\end{equation} 
This formulation treats all racial groups as equally important constituents of the urban system, measuring segregation as the mean level of isolation experienced by all groups in the city's mobility network. More details can be found in Supplementary Note 5.1.

\paragraph{Gravity-Based Segregation Index}

By comparing the gravity-based null model and the observed row-mormalized mixing matrix, we construct the gravity-based segregation matrix ($M_{gs}$): $M_{gs} = M_{r}-M_{g}$, which isolates mobility segregation patterns after rigorously accounting for both demographic distributions and spatial constraints. Whereas $M_{bs}$ only adjusts for demographic parity, $M_{gs}$ specifically controls for the fundamental tendency of mobility flows to decay with distance---a critical confounder since racial groups often cluster spatially. Each element $m^{gs}_{ij}$ represents the racially motivated deviation in mobility: positive values indicate flows exceeding expectations from spatial/demographic factors alone (suggesting active preference), while negative values reveal suppressed interactions attributable to racial barriers. By incorporating two scientifically grounded components: self-segregation term and cross-group mixing term, an analogous group-specific gravity-based segregation index for group $i$ is defined as:

\begin{equation}
    S_g^{i} = \underbrace{m^{gs}_{ii}}_{\text{self-segregation}} 
- \underbrace{\frac{1}{3}\sum_{i \neq j} m^{gs}_{ij}}_{\text{cross-group mixing}}
\end{equation}
in which $m^{gs}_{ii}, m^{gs}_{ij} \in M_{gs}$. Similar to the baseline index, the citywide gravity-based segregation index $S_g$ is defined as the average of all group-specific indices (Supplementary Note 5.2).

The segregation indices $ S_b^{i} $ and $ S_g^{i} $ exhibit similar theoretical ranges and converge to similar practical bounds under real-world constraints. Theoretically, $ S_b^{i}$ and $S_g^{i}$ have wider theoretical ranges of $[-2, 2]$, derived from extreme scenarios of flow redistribution. However, empirical analyses reveal that both indices typically fall within $[-1, 1]$ in practice (Supplement Note 5.2). This practical alignment enables direct comparison of the indices, though their interpretations differ.

\subsection{Homophily Gravity Model}

Building on the empirical motivation established in Section \ref{model_result}, we propose the Homophily Gravity Model (HGM), a generalized framework for quantifying how social affinity (e.g., racial, religious, or socioeconomic similarity) modulates spatial interactions. By introducing $s_{ij}$ quantifying dyadic social similarity, the Homophily Gravity Model takes the form:

\begin{equation}
    F_{ij} = \lambda \frac{P_{i}^\alpha P_{j}^\beta \cdot {s_{ij}^\gamma}}{d_{ij}^\delta}
\end{equation}

where $\lambda$ is a constant of proportionality, $F_{ij}$ represents the flow between $i$ and $j$ tracts, $P_{i}$ and $P_{j}$ are relative to the number of trips leaving the $i$ or the ones attracted by $j$, and $d_{ij}$ is the Euclidean distance between the centroids of $i$ and $j$ (Supplementary Note 6.1). HGM advances classical gravity models by explicitly embedding social dimensions (e.g., racial homophily) into spatial interaction rules. While this study focuses on racial homophily, the model's structure is extensible to other dimensions of social identity. For racial homophily in our study, we define $s_{ij} \equiv h_{ij}$ as the racial similarity parameter between tract $i$ and $j$, and construct $h_{ij}$ through two steps: a symmetric operation on the racial distributions of two tracts and an exponential transformation (Supplementary Note 6.2). Subsequently, the parameter $h_{ij}$ is incorporated into the HGM as a multiplicative factor, where its $\gamma$-exponentiated form ($h^\gamma$) captures how racial composition moderates destination attractiveness. Consistent with Section \ref{null_model}, we employ dynamically measured active populations to implement HGM. Finally, the HGM is formulated as:

\begin{equation}
F_{ij} = \lambda\ \frac{f_{i,out}^\alpha f_{j,in}^\beta h_{ij}^\gamma}{d_{ij}^\delta}
\end{equation}

where $\lambda$ is a constant of proportionality, $F_{ij}$ represents the flow between tracts $i$ and $j$, $f_{i,out}$ and $f_{j,in}$ are out-flow of CT $i$ and in-flow of CT $j$, and $d_{ij}$ is the Euclidean distance between the centroids of tracts $i$ and $j$. The exponents $\alpha$, $\beta$, $\gamma$, and $\delta$ were estimated through model fitting, with $\gamma$ specifically capturing the effect of racial homophily on mobility flows. To isolate the specific contribution of racial homophily, we contrast this enhanced specification with a Baseline Gravity Model (BGM) that excludes the $h_{ij}$ term\cite{barthelemy2011spatial,nie2023examining}. More details can be found in Supplementary Note 6.


\section*{Acknowledgement}
This work is supported by National Natural Science Foundation of China (Grant Nos. 72288101, 72271019).

\section*{Authorship Contribution Statement} \textbf{Wei-Peng NIE}: Conceptualization, Data curation, Formal analysis, Investigation, Methodology, Software, Writing – original draft, Writing review \& editing. \textbf{Ding-Tian Rong}: Validation, Visualization, Writing – review \& editing. \textbf{Xiao-Yong Yan}: Funding acquisition, Methodology, Supervision. \textbf{Tao Zhou}: Methodology, Supervision. \textbf{Zi-You Gao}: Funding acquisition, Methodology, Project administration, Supervision.

\section*{Competing interests}  
The authors declare no competing interests.

\newpage 

\begin{appendices}

\pagestyle{fancy} 
\setcounter{page}{1}  

\begin{center}
\vspace*{0.3cm} 
\Large\bfseries Supplementary Information
\vspace{0.3cm}
\end{center}
\addcontentsline{toc}{section}{Supplementary Information} 



\counterwithin{figure}{section}  
\renewcommand{\thefigure}{S\arabic{figure}} 

\counterwithin{table}{section}
\renewcommand{\thetable}{S\arabic{table}}


\renewcommand{\thesection}{\arabic{section}} 

\startcontents
\printcontents{}{1}{\section*{Supplementary Note}}
\vspace{1cm}

\section{Data Description and Processing}

\subsection{Human Mobility Data}

In our study, we utilize an anonymized human mobility dataset across New York City (NYC) for the period January 2019–April 2021. This dataset captures dynamic population movements during both pre-pandemic and pandemic phases. This dataset originates from mobile device location records in SafeGraph's Patterns database, containing anonymized GPS pings from mobile devices, subsequently processed and spatially aggregated by Kang et al.\cite{kang2020multiscale} through rigorous methodologies. Kang et al. clustered GPS pings of each user and mapped those clusters into $2,327$ census tracts (CTs) in NYC. For visitor flow calculation, they analyzed daily movement patterns by tracking clusters that appeared outside their home tract and count the visitor number from each home CT to other CTs. Finally, the accessible dataset provides daily origin-destination (OD) matrices containing tract GEOIDs, tract geographical coordinates,
date stamps, and visitor counts and is publicly available via GitHub (\url{https://github.com/GeoDS/COVID19USFlows?tab=readme-ov-file}). 

In this study, we aggregated these daily visitor flows into monthly totals, representing the cumulative number of visitors travelling between each census tract pair per month. In the pre-pandemic period, New York City averaged approximately 24.4 million monthly inter-tract visitor flows, establishing a baseline for our subsequent analysis of mobility segregation. Following pandemic onset, monthly visitor flows plummeted to a trough of just 11.3 million in April 2020 - a nearly $53\%$ reduction from baseline levels. Notably, the dataset captures unique visitor counts rather than trips. Compared to traditional trip-based metrics, this methodology better reflects the breadth (rather than frequency) of cross-neighborhood mobility, making it particularly suitable for studying segregation patterns where the primary interest lies in whether people visit different communities, not how often they do so.

\subsection{Racial Composition Data}

Demographic context was integrated using tract-level racial composition data from the 2020 Decennial Census (U.S. Census Bureau, \url{https://www.census.gov/}), which provides population proportions for White, Black/African American, Asian, Hispanic/Latino, and other racial groups of each Census Tract in NYC. All demographic attributes were then linked to mobility observations through tract-level GEOIDs to ensure spatial consistency across datasets. Although mobility data span 2019–2021, the 2020 Decennial Census serves as the demographic baseline due to the relative temporal stability of neighborhood-level racial composition. Prior studies and assessments using ACS estimates indicate that racial proportions at the census tract level exhibit only modest fluctuations year-over-year, particularly in dominant group shares\cite{us2020understanding}. Therefore, using static 2020 data is unlikely to bias segregation pattern analyses during the pandemic.
Nevertheless, the use of static 2020 racial composition data assumes negligible short-term demographic shifts, which may not fully capture localized changes driven by migration, displacement, or pandemic-induced residential mobility.

\subsection{Census Tract Selection Criteria}
\label{ct-select}

Our analysis focuses on 2,327 census tracts in New York City as defined by the 2020 U.S. Census. To ensure demographic interpretability and statistical robustness, we implemented a multi-stage selection protocol. First, tracts with populations below 100 residents were excluded from racial categorization, as small sample sizes induce substantial uncertainty in racial proportion estimates. In this operation, 101 CTs ($4.3\%$) are excluded due to populations below 100 residents. Second, we identified tracts with unambiguous racial dominance by requiring a single racial/ethnic group to constitute $\geq50\%$ of the population—a well-established threshold in neighborhood segregation studies\cite{wang2018urban}. The $\geq50\% $threshold ensures that the dominant racial/ethnic group not only holds plurality but also exceeds the combined share of all other groups, thus representing a socially salient community identity in segregation dynamics. This dual criterion yielded four mutually exclusive categories: majority-White (631 CTs), majority-Black (403 CTs), majority-Hispanic (394 CTs), and majority-Asian (141 CTs), totalling 1,569 CTs ($67.4\%$ of all CTs in NYC). Geospatially, majority-White tracts cluster in Staten Island and southern Brooklyn, while majority-Black tracts concentrate in central Brooklyn and southeast Queens. This spatial clustering reinforces the link between residential segregation and mobility patterns. The remaining CTs not meeting the $\geq 50\%$ threshold for any single racial group are excluded from our primary analysis. Excluding highly mixed tracts enhances the interpretability of cross-community mobility flows, as trips originating from mixed tracts can dilute directional exposure patterns. This selection approach enables examination of mobility patterns between racial communities while maintaining adequate sample sizes across all four major racial groups and minimizing demographic noise from highly mixed neighborhoods. 

To evaluate threshold sensitivity, we conducted robustness analyses using a stricter $\geq70\%$ majority criterion (Supplemental Note \ref{Sensitivity}). This alternative classification reduces tract counts to: White$=266$, Black$=204$, Hispanic$=124$, Asian$=40$ (total $n=634$, $27.2\%$ of tracts), but confirms core findings persist in unambiguously homogeneous neighborhoods. This sensitivity test evaluates whether segregation patterns hold in tracts with unambiguous racial homogeneity, mitigating potential misclassification bias.

\section{Temporal Phased Containment Measures in NYC}

New York City went through the following phases of containment measures during the COVID-19 pandemic\cite{huang2022lockdown}:

(1) \textbf{Initial Response Phase (March 2020)}

New York City swiftly implemented containment measures following its first confirmed COVID-19 case in early March 2020. After Governor Cuomo declared a state of emergency on March 7, a statewide stay-at-home order (PAUSE policy) took effect on March 22. This phase featured comprehensive non-essential business closures, strict stay-at-home requirements, and a ban on gatherings exceeding 10 people.  These early interventions established critical pandemic response infrastructure.

(2) \textbf{Strict Lockdown Phase (April-June 2020)}

As cases surged in April, New York City escalated to its most stringent containment measures. The extended NY PAUSE policy maintained closures of non-essential businesses and kept schools in remote learning mode. Essential venues like subways and supermarkets operated under strict capacity limits. These aggressive measures successfully curbed viral transmission by mid-June.

(3) \textbf{Phased Reopening (June-October 2020)}

New York City implemented a gradual four-phase reopening from June to July 2020 as COVID-19 cases decreased. The phased approach sequentially allowed different economic sectors to resume operations with capacity limits and safety protocols. During this period, the city simultaneously employed a "micro-cluster" strategy that imposed targeted restrictions in high-risk areas. Neighborhoods with elevated infection rates faced stricter limitations, including reduced business capacities and stronger mobility recommendations.

(4) \textbf{Winter Resurgence Response (November 2020-May 2021)}

Facing a seasonal resurgence, health authorities implemented a novel "micro-cluster" strategy in November 2020. Neighborhoods were classified into color-coded zones (Red/Orange/Yellow) based on infection rates, triggering tiered restrictions ranging from non-essential business closures in Red Zones to capacity-limited operations in Orange Zones. Concurrently, the December 2020 vaccine rollout prioritized healthcare workers and elderly residents before expanding to all adults by April 2021. This dual approach of targeted containment and mass vaccination prevented another full lockdown while controlling community transmission.

(5) \textbf{recovery and normalization phase (June 2021- March 2022)}

New York City lifted most capacity restrictions on July 1 2021 as vaccination rates increased. By March 2022, the city had discontinued its remaining mandates including indoor mask requirements and vaccine verification systems, marking the full transition to post-pandemic operations while maintaining disease surveillance infrastructure.

\section{Analyzing Mixing Pattern of Racialized Mobility Network}

\subsection{Racialized Mobility Network Construction}\label{net-con}

The racialized mobility network is constructed as a directed, weighted graph $G=(V,E)$ where nodes represent census tracts annotated by their dominant racial group ($\leq50\%$ population threshold) and edges capture monthly visitor flows between tracts, derived from SafeGraph mobility data. Each edge weight $w_{ij}$ reflects the visitor flow from tract $i$ to $j$. 
This approach of attributing characteristics to nodes within network architecture finds extensive application in the study of complex networks\cite{hu2019segregation,moro2021mobility}. The network can be partitioned into intra-group and inter-group components by aggregating tracts into four racial modules (White, Black, Hispanic, Asian), enabling comparative analysis of segregation patterns in mobility flows. This racial partitioning exhibits statistically significant community structure (modularity $Q = 0.20$, $p < 0.001$ via $10,000$ random network permutations), confirming that tracts preferentially exchange visitors within their own racial groups \cite{newman2004finding}. The modularity calculation accounts for the difference between observed intra-group flows and expected flows under random mixing, with the high $Q$ value indicating strong racial clustering in mobility patterns. Robustness checks confirm these findings persist across alternative edge weight definitions (raw counts, log-transformed flows) and community detection algorithms (Infomap), with temporal analysis showing stable modularity values ($0.18-0.22$) throughout January 2019-February 2020. The network construction methodology thereby captures both the intensity and racial structuring of urban mobility patterns.

\subsection{Mobility Mixing Matrix}

The mobility mixing matrix $M_{m}$ represents normalized flow probabilities from census tracts dominated by one racial group to tracts dominated by the same or different groups \cite{newman2003mixing}. Each tract is assigned to its majority racial group according to the classification described in Supplementary Note~\ref{ct-select}. We aggregate all outflows from nodes (tracts) dominated by racial group $i$ to nodes dominated by racial group $j$, and normalize by the total outflows from all tracts in the network (TotalOutflows). Formally, each element $m_{ij} \in M_{m}$ is defined as:
\begin{equation}
    m_{ij} = \text{Flows}(i \to j) / \text{TotalOutflows}
\end{equation}
where $i,j \in \{\text{White, Black, Hispanic, Asian}\}$ denote the four racial groups under study. The mixing matrix exhibits three fundamental characteristics: (1) its elements are globally normalized such that their sum satisfies $\sum_{i=1}^4\sum_{j=1}^4 m_{ij} = 1$; (2) it typically demonstrates directional asymmetry with $m_{ij} \neq m_{ji}$; (3) each element $m_{ij}$ carries a density interpretation representing the relative probability mass of all $i \rightarrow j$ flows within the complete mobility network.

To highlight patterns of segregation or integration, we distinguish between \emph{intra-group} probabilities ($m_{ii}$) and \emph{inter-group} probabilities ($m_{ij}$ with $i \neq j$). A high diagonal element indicates stronger within-group mobility preference, whereas larger off-diagonal elements signal more cross-group interactions. Temporal sequences of $M_{m}$ allow us to track how mobility segregation evolved across different phases of the COVID-19 pandemic.

\subsection{Calculating Assortativity Indices}

To quantify the structural mixing patterns encoded in the mobility mixing matrix, we introduce several assortativity indices that generalize concepts from network assortativity to racial mobility networks \cite{hu2019segregation,newman2003mixing}. These indices are used to quantify whether and to what extent links tend to connect nodes of the same type.

1. \textbf{Assortativity Coefficient ($r_{ac}$)}: Measures the preference for nodes to attach to others of the same type in a network \cite{newman2003mixing}. For a mixing matrix $M$ with elements $m_{ij}$ representing flows from group $i$ to $j$, $r_{ac}$ is defined as:

\begin{equation}
    r_{ac} = \frac{\sum_{i} m_{ii} - \sum_{i} a_i b_i}{1 - \sum_{i} a_i b_i}, \quad \text{where } a_i = \sum_j m_{ij}, \, b_i = \sum_i m_{ij}
\end{equation}

2. \textbf{E-I Index ($r_{ei}$)}: Quantifies the relative prevalence of inter-group versus intra-group links \cite{krackhardt1988informal}. For consistent interpretation across all indices (where positive values indicate assortative mixing), we invert the classical E-I Index by multiplying by -1, aligning its directionality with the other three metrics. For a mixing matrix $M$ with elements $m_{ij}$ representing flows from group $i$ to $j$, $r_{ei}$ is defined as:

\begin{equation}
    r_{ei} = -1\times\frac{\sum_{i} \sum_{i \neq j} m_{ij} - \sum_{i} m_{ii}}{\sum_{i} \sum_{j} m_{ij}}
\end{equation}

3. \textbf{Gupta–Anderson–May Index ($r_{gam}$)}:
Evaluates intra-group link concentration normalized by group count $K$\cite{gupta1989networks}. For a mixing matrix $M$ with elements $m_{ij}$ representing flows from group $i$ to $j$, $r_{gam}$ is defined as:

\begin{equation}
    r_{gam} = \frac{\sum_{i} f_{ii} - 1}{K - 1}, \quad \text{where } f_{ii} = \frac{m_{ii}}{\sum_j m_{ij}}
\end{equation}

4.  \textbf{Diagonality Index ($r_{di}$)}: Pearson correlation of matrix entries measuring diagonal concentration \cite{hilman2022socioeconomic}. For a mixing matrix $M$ with elements $m_{ij}$ representing flows from group $i$ to $j$, $r_{di}$ is defined as:

\begin{equation}
    r_{di} = \frac{\sum_{ij} ij \cdot m_{ij} - \left(\sum_{ij} i \cdot m_{ij}\right)\left(\sum_{ij} j \cdot m_{ij}\right)}{\sqrt{\sum_{ij} i^2 m_{ij} - \left(\sum_{ij} i m_{ij}\right)^2} \sqrt{\sum_{ij} j^2 m_{ij} - \left(\sum_{ij} j m_{ij}\right)^2}}
\end{equation}

All four indices range between $-1$ and $1$. A value of $1$ indicates perfect assortative mixing, corresponding to fully segregated mobility patterns, where mobility occurs exclusively within the same racial group (diagonal dominance in the mixing matrix). Conversely, $-1$ represents complete disassortative mixing, with flows existing only between different groups. A value near $0$ implies random, group-agnostic mobility patterns without segregation or preference.

\subsection{Row-normalized Mixing Matrix}

Conventional mixing matrices exhibit a critical methodological constraint: they inherently privilege majority racial groups in segregation analysis by conflating population distribution with behavioral patterns. This structural limitation arises because raw flow counts disproportionately reflect the mobility behaviors of numerically dominant groups—not due to stronger segregation tendencies, but simply because their larger spatial footprint generates more observable trips. In New York City's context, White residents' movements necessarily dominate the unnormalized matrix (representing $40.2\%$ of all majority tracts) while Asian residents' patterns ($8.9\%$ of tracts) become statistically marginalized, even if both groups exhibit identical propensity for within-group mobility.

Due to the unique nature of racial problem, we prioritize treating all racial groups equally when studying segregation patterns. To address the demographic imbalance in raw flow counts, we construct a row-normalized mixing matrix $M_{r}$ where each element $m^r_{ij}$ represents the proportion of outflows from racial group $i$ that terminate in group $j$, relative to the total outflows of group $i$ ($\text{TotalOutflows}(i)$). $m^r_{ij}$ is given by:

\begin{equation}
    m^r_{ij} = \text{Flows}(i \to j) / \text{TotalOutflows}(i)
\end{equation}
Row-normalized mixing matrix rescales all racial groups to equal footing—a single unit of outflow from a Black-majority tract carries identical analytical weight as one from a White-majority tract. By scaling each group's outflows to unity, this normalization removes population-size effects, permitting unbiased comparisons of segregation patterns of each race. Our normalization protocol operationalizes the principle of demographic equity in segregation research, ensuring that: (1) all racial groups contribute equally to segregation metrics regardless of population size, and (2) observed patterns reflect actual mobility preferences rather than distributional artifacts.

\section{Implementing Null Models}

The observed mixing matrix captures raw mobility flow proportions between racial groups, reflecting actual visitation patterns in the urban environment. However, these measured flows inherently conflate genuine segregation behaviors with various structural constraints --- including uneven population distributions, spatial clustering of racial groups, and the geographic arrangement of opportunities. To disentangle these confounding factors from true segregation signals, we construct null model mixing matrices that simulate how mobility would occur under idealized conditions. These theoretically-derived matrices serve as counterfactual benchmarks, revealing the expected flow patterns when racial preferences are absent and only structural factors (demographic composition and spatial organization) govern movement.

\subsection{Baseline Null Matrix}
\label{null_model}

The baseline null matrix $M_b$ is a special row-normalized mixing matrix assuming that racial mixing probabilities are strictly proportional to the geographic distribution of racial groups. Each element $c_{ij}$ in row $i$ (representing an origin racial group) is identical and defined as:

\begin{equation}
     m^b_{ij} = \frac{N_j}{\sum\limits_{k=1}^4 N_k}
\end{equation}
where $N_j$ is the count of tracts dominated by group $j$. 
For example, if Hispanic-majority tracts comprise $25\%$ of all tracts, we assume that they would receive $25\%$ of flows from every origin group. The matrix therefore serves as a critical reference point that captures expected mobility patterns under conditions of perfect demographic neutrality, where no racial group exhibits preferential attachment or avoidance behaviors, and where spatial arrangements impose no additional constraints on mobility beyond the basic distribution of racial groups across the urban landscape. This theoretical construct enables the isolation and quantification of deviations from this neutral baseline that may signal the presence of genuine segregation mechanisms in observed mobility patterns. This proportional mixing assumption creates a simplified baseline for comparing actual mobility patterns, helping identify significant deviations that may indicate segregation.

\subsection{Gravity-Based Null Matrix}

Despite recent challenges to the gravity model’s universal applicability in human mobility studies\cite{barthelemy2011spatial,simini2012universal}, it remains one of the most enduring and widely validated frameworks for modeling macro-scale spatial interactions. Originally inspired by Newtonian physics, the gravity formulation has been successfully adapted to explain a wide range of socio-spatial phenomena, including commuting flows, inter-city migration, international trade, and freight movements\cite{barthelemy2011spatial}. Its parsimonious structure—capturing the essential decay of interaction probability with increasing spatial separation while incorporating population or opportunity size—offers both interpretability and flexibility. In our context, its transparent parameterization allows us to establish a clear and theoretically grounded baseline for expected flows, against which we can systematically isolate deviations attributable to racial segregation effects\cite{yabe2025behaviour}.

Here, we construct a gravity-based null matrix $M_{g}$ by a null gravity model to incorporates spatial constraints to observed flows. In our implementation, we address key limitations of traditional gravity models by: (1) using dynamically measured active populations (unique visitor counts) as attractiveness proxies, which better reflect actual destination appeal than residential population; and (2) generating an ensemble of 100 synthetic networks to mitigate parameter uncertainty. This approach allows us to construct a robust null expectation of mobility flows that accounts for both demographic distributions and spatial constraints, forming an essential counterfactual for identifying racially motivated mobility biases. The equation of the null gravity model takes the following form:

\begin{equation}
    F_{ij} = \lambda\frac{f_{i,out}^\alpha f_{j,in}^\beta}{d_{ij}^\delta }
\end{equation}

where $f_{i,out}$ and $f_{j,in}$ is outflow of origin tract $i$ and inflow of destination tract $j$ respectively; $d_{ij}$ denotes the Euclidean distance between the centroids of tracts $i$ and $j$, and $\lambda$, $\alpha$, $\beta$, $\delta$ are parameters fitted via maximum likelihood estimation. While traditional gravity models often use residential population as a proxy for location attractiveness, we employ the dynamic active population - the count of unique visitors to or from a location - which better captures the actual human activity patterns and temporal variations in destination attractiveness \cite{barthelemy2011spatial, nie2023examining}. Using the fitted gravity model, we generate a synthetic flow network where each link $F _{ij}$ represents the expected flow under spatial and demographic constraints. To ensure the robustness and comparability of the gravity-based null network, we adopt the following construction procedure:

(1) Generate 100 synthetic flow networks using the optimized gravity model parameters

(2) Strictly preserve each tract's total outflows to match empirical observations

(3) Construct the gravity-based null matrix $M_{g}$ by taking the element-wise mean of row-normalized mixing matrices from all synthetic networks

In above procedure, we adopt a single-constrained gravity null that preserves each origin tract’s empirical total outflow (row sums) while leaving destination totals unconstrained. This choice deliberately removes the observed distribution of destination “capacity” from the benchmark—capacities that encode long-run agglomeration, land-use, and opportunity inequalities often correlated with neighborhood racial composition. By allowing destination inflows to be generated endogenously by distance decay and the overall trip supply, the null approximates an opportunity-neutral allocation: destinations do not inherit historically concentrated advantages (e.g., systematically larger inflows to central business districts). Residuals thus measure the distance- and demography-adjusted gap between realized mixing and a more equitable counterfactual, yielding an interpretable upper bound on segregation not attributable to destination capacity alone. Practically, the single constraint also improves numerical stability for sparse OD matrices and avoids fitting noise in observed inflow margins. This model accounts for both demographic distributions and the friction of distance, providing a more realistic counterfactual than the proportional baseline. Comparisons between actual row-normalized mixing matrices and these null models reveal whether observed segregation exceeds expectations from purely geographic/demographic factors.

\section{Segregation Index Derivations}

To quantify racial segregation in mobility patterns, we propose two classes of indices derived by comparing observed mixing matrices against null models:

\subsection{Baseline Segregation Index}
To evaluate segregation relative to demographic parity, we construct the baseline segregation matrix ($M_{bs}$) by subtracting the baseline null matrix ($M_b$) from the row-normalized mixing matrix ($M_r$):

\begin{equation}
    M_{bs} = M_{r}-M_{b}
\end{equation}
which quantifies how observed mixing deviates from a society where mobility probabilities strictly follow tract composition. The elements of the baseline segregation matrix $M_{bs}$ reveal meaningful patterns about racial mobility preferences. Positive values $m^{bs}_{ij}>0$ indicate that the observed flows from group $i$ to group $j$ exceed what would be expected under demographic parity, suggesting either preferential movement toward group $j$. Conversely, negative values $m^{bs}_{ij}<0$ signify suppressed mobility below demographic expectations, which may result from avoidance behaviors, institutional barriers, or socioeconomic constraints limiting cross-group interaction. Values near zero imply that mobility patterns align with what would be predicted solely by the racial composition of census tracts. The diagonal entries $m^{bs}_{ii}$ specifically measure the intensity of within-group segregation, reflecting self-segregation tendencies in co-racial mobility, which is the self-segregation signal. The off-diagonal elements $m^{bs}_{ij}$ ($i\neq j$) capture asymmetric dynamics in how different racial groups mixing spatially, which is the cross-group mixing signal. This matrix thus provides a nuanced, directional assessment of segregation that distinguishes between structural constraints of demography and active mobility preferences. For example, a positive $ m_{ii}\in M_{bs}$ indicates group $i$ exhibits stronger self-segregation than their CT proportion would predict. Based on the baseline segregation matrix, we propose group-specific segregation indices. The group-specific baseline segregation index for race $i$ consists of two components and is defined as:

\begin{equation}
    S_b^{i} = \underbrace{m^{bs}_{ii}}_{\text{self-segregation}} 
- \underbrace{\frac{1}{3}\sum_{i \neq j} m^{bs}_{ij}}_{\text{cross-group mixing}}
\end{equation}
in which the self-segregation item quantifies group $i$'s self-segregation degree (avoidance of out-group mobility), and the cross-group mixing term captures the other groups' visitation preferences toward group $i$. This index measures the net segregation for group $i$ by contrasting its diagonal excess with off-diagonal deficits. Thus, the segregation index captures the combined effect of group $i$'s tendency to self-isolate (reflected in the positive diagonal term) and the aggregate reluctance of other groups to visit group $i$ (captured by the negative cross-group term). This dual-term design operationalizes Schelling's theory of segregation dynamics\cite{1971Dynamic}. The city-wide baseline segregation index is defined as the average of all group-specific indices:

\begin{equation}
   S_b = \frac{1}{4}\sum_{i=1}^4 m^{bs}_{ii} 
- \frac{1}{12}\sum_{i \neq j} m^{bs}_{ij} = \frac{1}{4}\sum_{i=1}^4 S_b^{i}
\end{equation} 
This formulation treats all racial groups as equally important constituents of the urban system, measuring segregation as the mean level of isolation experienced by all groups in the city's mobility network. 

\subsection{Gravity-Based Segregation Index}

We repeat the analysis utilizing the gravity-based null model, and construct the gravity-based segregation matrix ($M_{gs}$):  

\begin{equation}
    M_{gs} = M_{r}-M_{g}
\end{equation}
Whereas $M_{bs}$ only adjusts for demographic parity, this residual matrix $M_{gs}$ isolates segregation attributable to racial factors after accounting for demographic parity and distance decay. Each element $m^{gs}_{ij}$ represents the racially motivated deviation in mobility: positive values indicate flows exceeding expectations from spatial/demographic factors alone (suggesting active preference), while negative values reveal suppressed interactions attributable to racial barriers. By incorporating two scientifically grounded components: self-segregation term and cross-group mixing term, analogous group-specific gravity-based segregation index for group $i$ is defined as:

\begin{equation}
    S_g^{i} = \underbrace{m^{gs}_{ii}}_{\text{self-segregation}} 
- \underbrace{\frac{1}{3}\sum_{i \neq j} m^{gs}_{ij}}_{\text{cross-group mixing}}
\end{equation}
in which $m^{gs}_{ii}, m^{gs}_{ij} \in M_{gs}$. The city-wide gravity-based segregation index is defined as:
\begin{equation}
   S_g = \frac{1}{4}\sum_{i=1}^4 m^{gs}_{ii} 
- \frac{1}{12}\sum_{i \neq j} m^{gs}_{ij} = \frac{1}{4}\sum_{i=1}^4 S_g^{i}
\end{equation} 

The segregation indices $ S_b^{i} $ and $ S_g^{i} $ exhibit similar theoretical ranges and converge to similar practical bounds under real-world constraints. Theoretically, $ S_b^{i} $  and $S_g^{i} $  have wider theoretical ranges of $[-2, 2]$, derived from extreme scenarios of flow redistribution. However, empirical analyses reveal that both indices typically fall within $[-1, 1]$ in practice, as real-world mobility patterns are constrained by: (1) row-normalization preserving total outflows, (2) geographic friction limiting extreme flow concentration, and (3) balanced racial distributions in urban settings. This practical alignment enables direct comparison of the indices, though their interpretations differ: $ S_b^{i}$ isolates demographic effects, while $ S_g^{i}$ captures residual segregation after accounting for both demography and spatial structure. For robust comparison, we recommend contextualizing values relative to their empirical distributions rather than  theoretical extremes.

\section{Racial Homophily Modeling and Analysis}

\subsection{Homophily Gravity Model}

Existing research has well established that spatial interactions are shaped by a complex interplay of geographical constraints and social dynamics. Beyond the fundamental effects of physical distance and population distribution, mobility patterns consistently reflect underlying socio-demographic preferences, including racial, ethnic, and socioeconomic homophily \cite{hu2019segregation,moro2021mobility,yabe2025behaviour}. Traditional spatial interaction models, while successfully characterizing the distance-decay effect and mass-attraction principles \cite{barthelemy2011spatial,tobler1970computer}, often fail to explicitly account for these critical social dimensions of mobility behavior. This theoretical gap significantly limits our ability to decode the segregation mechanisms inherent in urban movement networks, especially in diverse metropolitan contexts where multiple social identities intersect. Our empirical findings underscore the urgent need to develop next-generation interaction models that formally integrate social affinity parameters as essential components of mobility impedance. Such advancement becomes particularly crucial for analyzing segregated urban systems, where layered social divisions interact with spatial separation to reinforce mobility barriers and network fragmentation.

Building on the empirical motivation established in Section 2.5, we propose the Homophily Gravity Model (HGM), a generalized framework for quantifying how social affinity (e.g., racial, religious, or socioeconomic similarity) modulates spatial interactions. By introducing $s_{ij}$  quantifying dyadic social similarity, the Homophily Gravity Model takes the form:

\begin{equation}
    F_{ij} = \lambda \frac{P_{i}^\alpha P_{j}^\beta \cdot {s_{ij}^\gamma}}{d_{ij}^\delta}
\end{equation}

where $\lambda$ is a constant of proportionality, $F_{ij}$ represents the flow between $i$ and $j$ tracts, $P_{i}$ and $P_{j}$ are relative to the number of trips leaving the $i$ or the ones attracted by $j$, and $d_{ij}$ is the Euclidean distance between the centroids of $i$ and $j$. HGM advances classical gravity models by explicitly embedding social dimensions (e.g., racial homophily) into spatial interaction rules, addressing long-standing critiques of purely geographic mobility models \cite{simini2012universal}. This framework bridges Schelling-style social dynamics with spatial interaction theory, offering a unified tool to dissect segregation mechanisms in mobility networks\cite{1971Dynamic}. While this study focuses on racial homophily, the model's structure is extensible to other dimensions of social identity.

\subsection{Homophily Gravity Model for Racial Homophily}

For racial homophily in our study, we define $s_{ij} \equiv h_{ij}$ as the racial similarity parameter. We construct $h_{ij}$ through two conceptual stages: First, the primary racial similarity score $(h'_{ij})$ is computed through a symmetric operation on the racial distributions of two tracts. Formally, for tracts $i$ and $j$ with racial population proportion vectors $r_i$ and $r_j$, $h'_{ij}$ is given by:

\begin{equation}
    h'_{ij} = \frac{1}{2} \times \{ max(r_i) \times r_j[argmax(r_i)] + max(r_j) \times r_i[argmax(r_j)] \}
\end{equation}

where $max(\cdot)$ extracts the dominant racial proportion and $argmax(\cdot)$ identifies the corresponding racial group. This symmetric design ensures maximal values when both tracts share the same dominant racial group with high proportions. For instance, if tract $i$ has $60\%$ White and $10\%$ Black residents while tract $j$ has $20\%$ White and $50\%$ Black residents, the computation would be $h'_{ij} = (0.6 \times 0.2 + 0.5\times 0.1)/2 = 0.085$. Second, to address the bounded nature of $h'_{ij} \in [0,1]$ and amplify sensitivity to segregation effects, we apply an exponential transformation $h_{ij} = e^{h'_{ij}}$, producing values in $[1,e]$ that better differentiate between integrated and segregated neighborhoods. The transformation reflects the expected nonlinear relationship between racial similarity and mobility behavior. The complete derivation maintains interpretability while addressing the empirical challenges of modelling segregation effects in population flows. 

Subsequently, the parameter $h_{ij}$ is incorporated into the HGM as a multiplicative factor, where its $\gamma$-exponentiated form ($h^\gamma$) captures how racial composition moderates destination attractiveness. Consistent with Section \ref{null_model}, we employ dynamically measured active populations to implement HGM. Finally, the HGM is formulated as:

\begin{equation}
F_{ij} = \lambda\ \frac{f_{i,out}^\alpha f_{j,in}^\beta h_{ij}^\gamma}{d_{ij}^\delta}
\end{equation}

where $\lambda$ is a constant of proportionality, $F_{ij}$ represents the flow between tracts $i$ and $j$, $f_{i,out}$ and $f_{j,in}$ are out-flow of CT $i$ and in-flow of CT $j$, and $d_{ij}$ is the Euclidean distance between the centroids of tracts $i$ and $j$. The exponents $\alpha$, $\beta$, $\gamma$, and $\delta$ were estimated through model fitting, with $\gamma$ specifically capturing the effect of racial homophily on mobility flows. To isolate the specific contribution of racial homophily, we contrast this enhanced specification with a Baseline Gravity Model (BGM) that excludes the $h_{ij}$ term\cite{barthelemy2011spatial,nie2023examining}, which is written as:
\begin{equation}
F_{ij}^{\text{baseline}} = \lambda\ \frac{f_{i,out}^\alpha f_{j,in}^\beta}{d_{ij}^\delta}
\end{equation}

\subsection{Model Calibration and Short-Distance Trips Analysis}
\label{Model—Calibration}

Initial application of the Homophily Gravity Model (HGM) and the Baseline Gravity Model (BGM) to the complete dataset (including all trip distances) yielded limited improvement over the baseline specification, with adjusted $R^2$ increasing marginally from 0.562 to 0.578 (see \hyperref[tab_full_data]{Table S1}). This modest enhancement stands in stark contrast to the more substantial gains observed when focusing on longer-distance mobility patterns, consistent with our prior findings that racial homophily effects are substantially attenuated in hyperlocal trips ($<1$ km) where trip generation is dominated by localized errands, social calls, and immediate-neighborhood amenities rather than socio-spatial sorting processes.

Building upon established evidence that sub-kilometer movements operate under fundamentally distinct mechanical regimes—characterized by habitual routines and micro-scale built environment factors rather than macro-scale gravitational principles—we subsequently excluded these short-distance flows to isolate the core dynamics of race-modulated spatial interaction. This methodological refinement, aligned with best practices in urban mobility modeling, enabled more precise quantification of gravitational parameters while controlling for confounding behavioral noise inherent to hyperlocal movement patterns.
This behavioral pattern aligns with established urban mobility research demonstrating that sub-kilometer trips operate under distinct mechanical regimes dominated by micro-scale built environment factors (sidewalk connectivity, land-use mix) and habitual routines rather than macro-scale gravitational principles, as evidenced through multi-country walkability studies\cite{van2009neighbourhood,wang2019neighbourhood}. Finally, the 1 km threshold retains $76\%$ of inter-tract flows, while it filters out micro-scale movements governed by pedestrian accessibility and immediate proximity-driven convenience, and preserves the core distance decay dynamics central to our analysis. By excluding hyperlocal trips, the analysis foregrounds cross-neighborhood mobility where racial homophily is more likely to manifest in destination choice, enabling cleaner identification of segregation-related deviations from baseline gravity predictions.

\setlength{\tabcolsep}{20pt} 
\begin{table}[h!]
  \centering
  \caption{Gravity Model Specifications (November 2019 - February 2020)}
  \label{tab_full_data}
  \begin{tabular}{lcc}
     \toprule
      \textbf{Parameter} & \textbf{Baseline (w/o $h_{ij}$)} & \textbf{Full (w/ $h_{ij}$)} \\
  \midrule
    \multicolumn{3}{l}{\textbf{Conventional Parameters}} \\
    $\lambda$ & 31.522*** & 18.334*** \\
              & (0.110)   & (0.074)   \\
    $\alpha$  & 1.456***  & 1.477***  \\
              & (0.002)   & (0.002)   \\
    $\beta$   & 0.793***  & 0.830***  \\
              & (0.001)   & (0.001)   \\
    $\delta$  & 1.225***  & 1.145***  \\
              & (0.001)   & (0.001)   \\
    \midrule
    \multicolumn{3}{l}{\textbf{Racial Parameter}} \\
    $\gamma$  & \multicolumn{1}{c}{--} & 1.119*** \\
              &                       & (0.006)  \\
    \midrule
    Observations & \multicolumn{2}{c}{950,689} \\
    $R^2$       & 0.5616  & 0.5779  \\
    Adjusted $R^2$ & 0.5616 & 0.5779 \\
    \bottomrule
     \multicolumn{3}{l}{\footnotesize \textit{Notes: *$p < 0.1$; **$p < 0.01$; ***$p < 0.001$.}}
  \end{tabular}
\end{table}

\section{Sensitivity and Robustness Analysis}\label{RSA}

\subsection{Extended Monthly Results}\label{apA}

The main text presents February 2020 as a representative pre-pandemic case study, but our findings generalize robustly across the entire observation period from January 2019 through April 2021. For completeness, We provide extended examples for additional months beyond the single monthly case presented in the main text. The patterns demonstrate consistency with the primary example. List of Tables and Figures:

\hyperref[tab:tableA2]{\textbf{Table S2-S5}}: Mixing matrix ($M_m$), row-normalized mixing matrix ($M_r$), baseline segregation matrix ($M_b$), and gravity-based segregation matrix ($M_g$) for May 2020.

\hyperref[tab:tableA6]{\textbf{Table S6-S9}}: Mixing matrix ($M_m$), row-normalized mixing matrix ($M_r$), baseline segregation matrix ($M_b$), and gravity-based segregation matrix ($M_g$) for December 2020.

\hyperref[fig:fig-si-expose]{\textbf{Fig.S1}}: Violin plots of exposure rates by racial group for January 2019, July 2019, May 2020 and December 2020.

\hyperref[fig:si-distance]{\textbf{Fig.S2}}: Average flow proportion $\bar R$ decays with $d_{ij}$ in May 2020 and December 2020

\hyperref[tab:A10]{\textbf{Table S10}}: Gravity Model Specifications for July 2019- October 2019.

\hyperref[tab:A11]{\textbf{Table S11}}: Gravity Model Specifications for July 2020 - October 2020.

\subsection{Sensitivity Analysis}\label{Sensitivity}

To ensure the robustness of our findings, we conduct a sensitivity analysis by repeating key experiments with a stricter threshold ($\geq 70\%$ majority-group CTs). This validation step tests whether observed segregation patterns persist when examining tracts with stronger racial demographic homogeneity, addressing potential biases introduced by the initial $\geq 50\%$ cutoff. The resulting sample comprised 266 White-majority, 204 Black-majority, 124 Hispanic-majority, and 40 Asian-majority census tracts, totaling 634, which accounts for approximately $27.2\%$ of all census tracts (2,327). In segregation research, increasing the majority threshold from $50\%$ to $70\%$ constitutes a shift from identifying “numerical majority” tracts to isolating “super-majority” or “homogeneous” tracts. This stricter criterion minimizes demographic noise from racially mixed areas, thereby magnifying the underlying social preference signals while controlling for compositional noise. It's worth noting that such “super-majority” tracts may yield noisier estimates due to their relatively small number and more scattered spatial distribution, which can weaken statistical stability and spatial representativeness.

The mixing matrix ($M_m$), row-normalized mixing matrix ($M_r$), baseline segregation matrix ($M_b$), and gravity-based segregation matrix ($M_g$) for February 2020 is shown in \hyperref[tab:tableA12]{Table S12-A15}.
From January 2019 to February 2020 before the pandemic, the average value of $r_{ac}$, $r_{ei}$, $r_{gam}$, $r_{d}$ is $0.7424\pm0.0043$,  $0.6649\pm0.0062$, $0.7056\pm0.0036$, $0.7125\pm0.0047$ (mean $\pm$ s.e.m.). The average value of $S_b^i$ is  $0.7382\pm0.0038$, $0.7365\pm0.0033$, $0.6965\pm0.0041$, $0.65033\pm0.0046$ (mean $\pm$ s.e.m.) for
White, Black, Hispanic, and Asian respectively, and $S_b$ is $0.7056\pm0.0036$. The average value of $S_g^i$ is $0.2543\pm0.0029$, $0.2270\pm0.0027$, $0.1507\pm0.0052$, $0.35780\pm0.0027$ (mean $\pm $s.e.m.) for
White, Black, Hispanic, and Asian respectively, and $S_g$ is $0.2478\pm0.0032$ (see \hyperref[fig:fig_si_70_seg]{Fig.S3 a-c}). All above indices showed higher values under the $70\%$ threshold compared to the $50\%$ threshold, indicating intensified racial isolation when analyzing tracts with stronger demographic homogeneity. Additionally, under the $70\%$ threshold, these metrics similarly exhibited pandemic-induced intensification patterns consistent with the $50\%$ threshold results (see \hyperref[fig:fig_si_70_seg]{Fig.S3 d-f}). The results of exposed rate distribution and percent change over time are shown in \hyperref[fig:fig_si_exposed_70]{Fig.S4}, which are similar to the results under the $50\%$ threshold. 

As illustrated in \hyperref[fig:si_distance_70]{Fig.S5}, average flow proportion $\bar R$ also decays with $d_{ij}$ with a slope $\bar R \propto d_{ij}^{\eta}$ ($\eta \in[-1.2,-1.1]$) under the $70\%$ threshold. Though network size reduction introduces irregular fluctuations in the curves, the decay patterns of intra-racial flows predominantly remain above other curves for White, Black, and Asian communities, indicating persistent spatial homophily. Hispanic communities absent the pronounced elevation in same-race flow curves. Despite the sharp reduction in network size (from 1,569 to 634 CTs) under the $70\%$ threshold—which inherently increases spatial dispersion and reduces sampling density—the gravity model with racial homophily ($h_{ij}$) consistently outperforms the baseline version. As evidenced in \hyperref[tab:model_2019_70]{Table S16} and \hyperref[tab:model_2020_70]{A17}, the inclusion of $h_{ij}$ elevates adjusted $R^2$ in two periods, demonstrating that racial alignment captures latent mobility preferences even in this sparser network. This improvement underscores the model's robustness to spatial sparsity: The predictive power of $h_{ij}$ persists despite fragmented CT distributions, suggesting its role transcends pure geographic proximity. 

\newpage

\newcolumntype{C}{>{\centering\arraybackslash}p{1.2cm}} 
\renewcommand{\arraystretch}{0.9} 

\setcounter{table}{1}
\begin{center}
\captionof{table}{ Mixing Matrix ($M_m$) for May 2020} \label{tab:tableA2}
\vspace{0.2cm} 
\begin{tabular}{@{} l | C C C C @{}}
\toprule
 & \textbf{White} & \textbf{Black} & \textbf{Hispanic} & \textbf{Asian} \\
\midrule
\textbf{White}  & 0.2685 & 0.0202 & 0.0291 & 0.0154 \\
\textbf{Black}   & 0.0474 & 0.1823 & 0.0377 & 0.0058 \\
\textbf{Hispanic}& 0.0566 & 0.0293 & 0.2330 & 0.0094 \\
\textbf{Asian}   & 0.0191 & 0.0040 & 0.0084 & 0.0338 \\
\bottomrule
\end{tabular}
\end{center}

\vspace{0.5cm} 

\begin{center}
\captionof{table}{ Row-Normalized Mixing Matrix ($M_r$) for May 2020} \label{tab:tableA3}
\vspace{0.2cm}
\begin{tabular}{@{} l | C C C C @{}}
\toprule
 & \textbf{White} & \textbf{Black} & \textbf{Hispanic} & \textbf{Asian} \\
\midrule
\textbf{White}  & 0.8058 & 0.0606 & 0.0873 & 0.0464 \\
\textbf{Black}   & 0.1733 & 0.6673 & 0.1382 & 0.0212 \\
\textbf{Hispanic}& 0.1725 & 0.0891 & 0.7097 & 0.0288 \\
\textbf{Asian}   & 0.2923 & 0.0613 & 0.1284 & 0.5181 \\
\bottomrule
\end{tabular}
\end{center}

\vspace{0.5cm}

\begin{center}
\captionof{table}{Baseline Segregation Matrix ($M_{bs}$) for May 2020} \label{tab:tableA4}
\vspace{0.2cm}
\begin{tabular}{@{} l | C C C C @{}}
\toprule
 & \textbf{White} & \textbf{Black} & \textbf{Hispanic} & \textbf{Asian} \\
\midrule
\textbf{White}  &  0.4036 & -0.1963 & -0.1638 & -0.0435 \\
\textbf{Black}   & -0.2288 &  0.4104 & -0.1130 & -0.0686 \\
\textbf{Hispanic}& -0.2297 & -0.1678 &  0.4586 & -0.0611 \\
\textbf{Asian}   & -0.1099 & -0.1956 & -0.1227 &  0.4283 \\
\bottomrule
\end{tabular}
\end{center}

\vspace{0.5cm}

\begin{center}
\captionof{table}{Gravity-Based Segregation Matrix ($M_{gs}$) for May 2020} \label{tab:tableA5}
\vspace{0.2cm}
\begin{tabular}{@{} l | C C C C @{}}
\toprule
 & \textbf{White} & \textbf{Black} & \textbf{Hispanic} & \textbf{Asian} \\
\midrule
\textbf{White}  &  0.2245 & -0.1258 & -0.0848 & -0.0138 \\
\textbf{Black}   & -0.0608 &  0.1347 & -0.0512 & -0.0227 \\
\textbf{Hispanic}& -0.0096 & -0.0450 &  0.0675 & -0.0129 \\
\textbf{Asian}   & -0.0311 & -0.1255 & -0.1112 &  0.2678 \\
\bottomrule
\end{tabular}
\end{center}

\vspace{0.5cm}

\begin{center}
\captionof{table}{ Mixing Matrix ($M_m$) for December 2020} \label{tab:tableA6}
\vspace{0.2cm}
\begin{tabular}{@{} l | C C C C @{}}
\toprule
 & \textbf{White} & \textbf{Black} & \textbf{Hispanic} & \textbf{Asian} \\
\midrule
\textbf{White}  & 0.3479 & 0.0225 & 0.0333 & 0.0209 \\
\textbf{Black}   & 0.0449 & 0.1385 & 0.0277 & 0.0056 \\
\textbf{Hispanic}& 0.0581 & 0.0238 & 0.1865 & 0.0112 \\
\textbf{Asian}   & 0.0264 & 0.0042 & 0.0088 & 0.0397 \\
\bottomrule
\end{tabular}
\end{center}

\vspace{0.5cm}

\begin{center}
\captionof{table}{ Row-Normalized Mixing Matrix ($M_r$) for December 2020} \label{tab:tableA7}
\vspace{0.2cm}
\begin{tabular}{@{} l | C C C C @{}}
\toprule
 & \textbf{White} & \textbf{Black} & \textbf{Hispanic} & \textbf{Asian} \\
\midrule
\textbf{White}  & 0.8194 & 0.0529 & 0.0784 & 0.0493 \\
\textbf{Black}   & 0.2071 & 0.6393 & 0.1277 & 0.0259 \\
\textbf{Hispanic}& 0.2079 & 0.0851 & 0.6668 & 0.0402 \\
\textbf{Asian}   & 0.3337 & 0.0526 & 0.1116 & 0.5020 \\
\bottomrule
\end{tabular}
\end{center}

\vspace{0.5cm}

\begin{center}
\captionof{table}{ Baseline Segregation Matrix ($M_{bs}$) for December 2020} \label{tab:tableA8}
\vspace{0.2cm}
\begin{tabular}{@{} l | C C C C @{}}
\toprule
 & \textbf{White} & \textbf{Black} & \textbf{Hispanic} & \textbf{Asian} \\
\midrule
\textbf{White}  &  0.4172 & -0.2040 & -0.1727 & -0.0405 \\
\textbf{Black}   & -0.1950 &  0.3824 & -0.1234 & -0.0640 \\
\textbf{Hispanic}& -0.1943 & -0.1718 &  0.4157 & -0.0497 \\
\textbf{Asian}   & -0.0685 & -0.2042 & -0.1395 &  0.4122 \\
\bottomrule
\end{tabular}
\end{center}

\vspace{0.5cm}

\begin{center}
\captionof{table}{ Gravity-Based Segregation Matrix ($M_{gs}$) for December 2020} \label{tab:tableA9}
\vspace{0.2cm}
\begin{tabular}{@{} l | C C C C @{}}
\toprule
 & \textbf{White} & \textbf{Black} & \textbf{Hispanic} & \textbf{Asian} \\
\midrule
\textbf{White} & 0.1850 & -0.1009 & -0.0693 & -0.0149 \\
\textbf{Black} & -0.0733 & 0.1470 & -0.0477 & -0.0260 \\
\textbf{Hispanic}& -0.0212 & -0.0410 & 0.0742 & -0.0120 \\
\textbf{Asian} & -0.0265 & -0.1058 & -0.1003 & 0.2326 \\
\bottomrule
\end{tabular}
\end{center}


\setcounter{figure}{0}
\begin{figure}[H]
    \centering 
    \includegraphics[width=15cm,trim=0 0 0 10,clip]{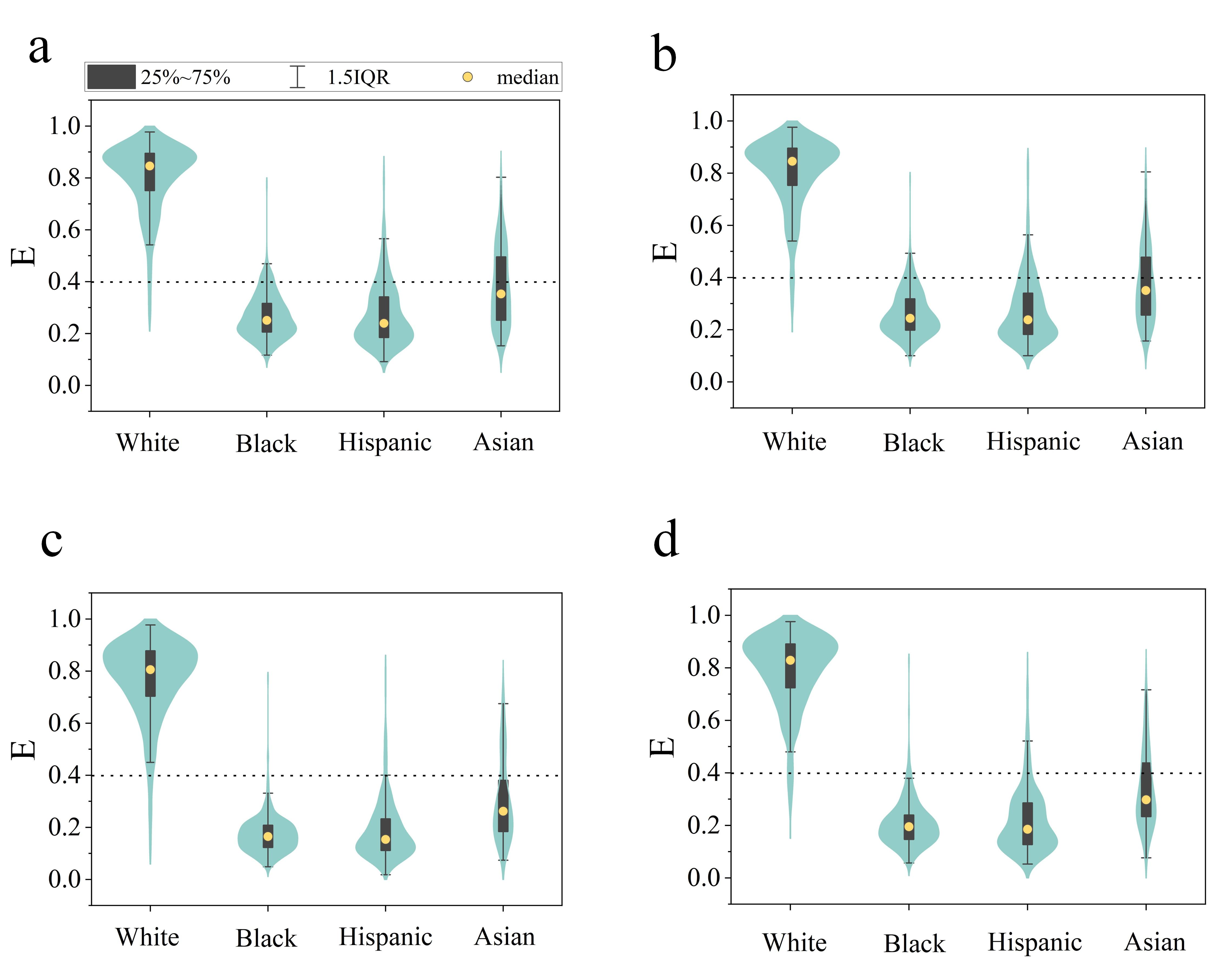}
    \caption{\textbf{a-d} Violin plots of exposure rates by racial group for January and July 2019, May and December 2020. Compared to Black, Hispanic, and Asian-majority CTs, White-majority CTs reveal significantly higher mean exposure in different months.}
	\label{fig:fig-si-expose}
\end{figure}

\begin{figure}[H]
    \centering 
    \includegraphics[width=13cm,trim=0 0 0 10,clip]{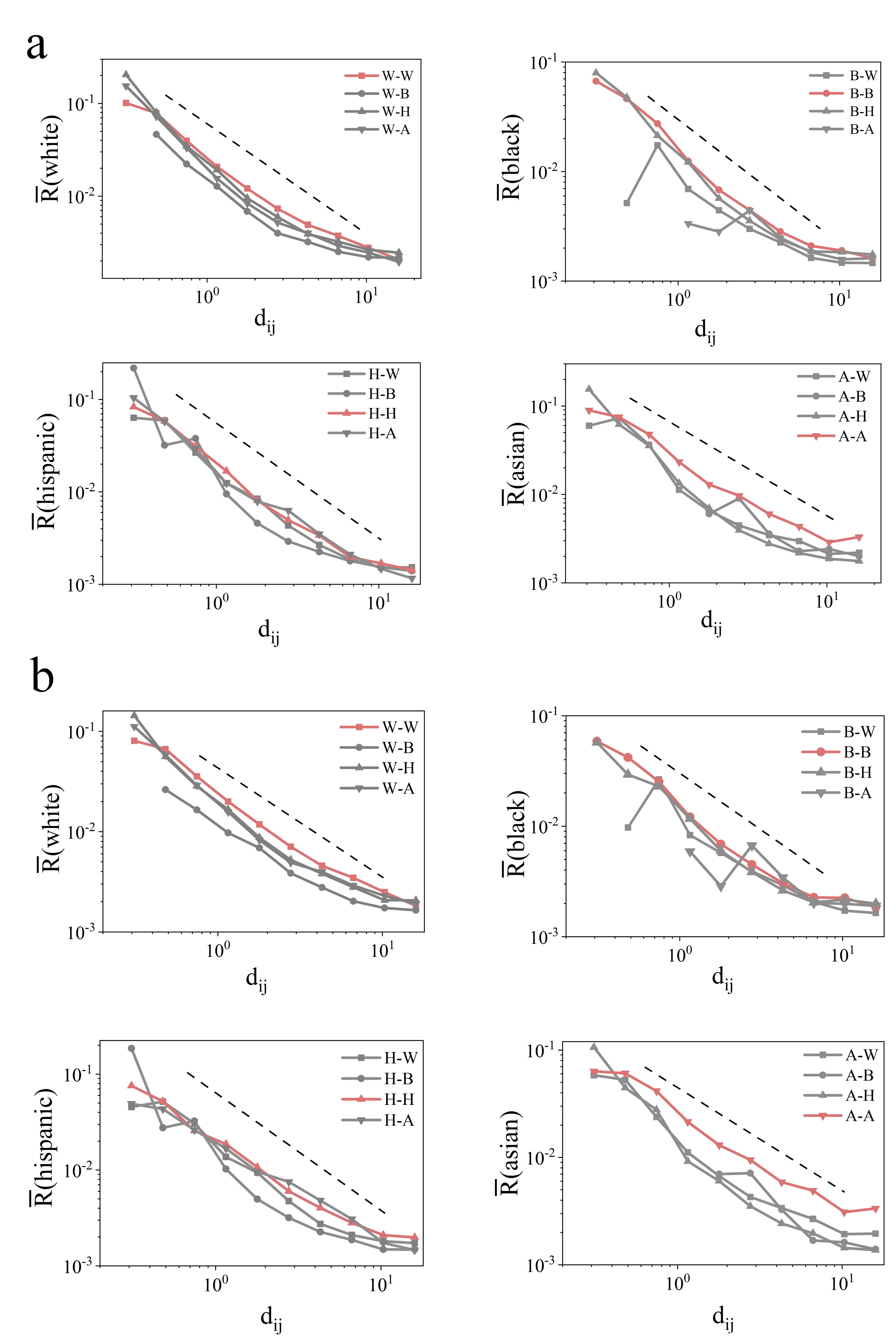}
     \caption{\textbf{a-b} Average flow proportion $\bar R$ decays with $d_{ij}$ with a slope $\bar R \propto d_{ij}^{\eta}$ in May 2020 and December 2020, in which $\eta \in[-1.2,-1.1]$. The curve representing flows between the same racial groups predominantly resided above the curves for flows between different groups across most distances analyzed. For Hispanic communities, the same-race flow curve did not maintain a significantly higher position relative to other curves.}
	\label{fig:si-distance}
\end{figure}

\begin{table}[h!]
  \centering
  \caption{Gravity Model Specifications (July 2019 - October 2019)}
  \label{tab:A10}
  \setlength{\tabcolsep}{25pt} 
  \begin{tabular}{lcc}
     \toprule
      \textbf{Parameter} & \textbf{Baseline (w/o $h_{ij}$)} & \textbf{Full (w/ $h_{ij}$)} \\
  \midrule
    \multicolumn{3}{l}{\textbf{Conventional Parameters}} \\
    $\lambda$ & 14.174***  & 3.200***   \\
              & (0.128)    & (0.036)    \\
    $\alpha$  & 1.344***   & 1.478***   \\
              & (0.004)    & (0.004)    \\
    $\beta$   & 0.993***   & 1.050***   \\
              & (0.003)    & (0.003)    \\
    $\delta$  & 1.116***   & 0.845***   \\
              & (0.003)    & (0.003)    \\
    \midrule
    \multicolumn{3}{l}{\textbf{Racial Parameter}} \\
    $\gamma$  & \multicolumn{1}{c}{--} & 2.021*** \\
              &                       & (0.013)  \\
    \midrule
    Observations & \multicolumn{2}{c}{1,014,738} \\
    $R^2$       & 0.273  & 0.311  \\
    Adjusted $R^2$ & 0.273 & 0.311 \\
    \bottomrule
     \multicolumn{3}{l}{\footnotesize \textit{Notes: *$p < 0.1$; **$p < 0.01$; ***$p < 0.001$.}}
  \end{tabular}
\end{table}

\vspace{1.5cm}

\begin{table}[h!]
  \centering
  \caption{Gravity Model Specifications (July 2020 - October 2020)}
  \label{tab:A11}
  \setlength{\tabcolsep}{25pt} 
  \begin{tabular}{lcc}
     \toprule
      \textbf{Parameter} & \textbf{Baseline (w/o $h_{ij}$)} & \textbf{Full (w/ $h_{ij}$)} \\
  \midrule
    \multicolumn{3}{l}{\textbf{Conventional Parameters}} \\
    $\lambda$ & 26.592*** & 12.393*** \\
              & (0.112)   & (0.066)   \\
    $\alpha$  & 1.849***  & 1.651***  \\
              & (0.003)   & (0.002)   \\
    $\beta$   & 1.168***  & 1.191***  \\
              & (0.001)   & (0.001)   \\
    $\delta$  & 1.206***  & 1.066***  \\
              & (0.002)   & (0.002)   \\
    \midrule
    \multicolumn{3}{l}{\textbf{Racial Parameter}} \\
    $\gamma$  & \multicolumn{1}{c}{--} & 2.007*** \\
              &                       & (0.008)  \\
    \midrule
    Observations & \multicolumn{2}{c}{770,911} \\
    $R^2$       & 0.452  & 0.494  \\
    Adjusted $R^2$ & 0.452 & 0.494 \\
    \bottomrule
     \multicolumn{3}{l}{\footnotesize \textit{Notes: *$p < 0.1$; **$p < 0.01$; ***$p < 0.001$.}}
  \end{tabular}
\end{table}

\clearpage

\begin{center}
\captionof{table}{Mixing Matrix ($M_m$) for February 2020 under the \\ 70\% threshold} \label{tab:tableA12}
\vspace{0.1cm}
\begin{tabular}{@{} l | C C C C @{}}
\toprule
 & \textbf{White} & \textbf{Black} & \textbf{Hispanic} & \textbf{Asian} \\
\midrule
\textbf{White} & 0.4203 & 0.0122 & 0.0095 & 0.0062 \\
\textbf{Black} & 0.0486 & 0.2508 & 0.0149 & 0.0046 \\
\textbf{Hispanic}& 0.0346 & 0.0095 & 0.1510 & 0.0039 \\
\textbf{Asian} & 0.0076 & 0.0013 & 0.0020 & 0.0230 \\
\bottomrule
\end{tabular}
\end{center}

\vspace{0.2cm}

\begin{center}
\captionof{table}{ Row-Normalized Mixing Matrix ($M_r$) for February \\ 2020 under the 70\% threshold} \label{tab:tableA13}
\vspace{0.1cm}
\begin{tabular}{@{} l | C C C C @{}}
\toprule
 & \textbf{White} & \textbf{Black} & \textbf{Hispanic} & \textbf{Asian} \\
\midrule
\textbf{White} & 0.9380 & 0.0272 & 0.0211 & 0.0138 \\
\textbf{Black} & 0.1522 & 0.7864 & 0.0468 & 0.0146 \\
\textbf{Hispanic}& 0.1737 & 0.0479 & 0.7587 & 0.0197 \\
\textbf{Asian} & 0.2246 & 0.0390 & 0.0587 & 0.6777 \\
\bottomrule
\end{tabular}
\end{center}

\vspace{0.2cm}

\begin{center}
\captionof{table}{ Baseline Segregation Matrix ($M_{bs}$) for February \\ 2020 under the 70\% threshold} \label{tab:tableA14}
\vspace{0.1cm}
\begin{tabular}{@{} l | C C C C @{}}
\toprule
 & \textbf{White} & \textbf{Black} & \textbf{Hispanic} & \textbf{Asian} \\
\midrule
\textbf{White} & 0.5358 & -0.2297 & -0.2300 & -0.0761 \\
\textbf{Black} & -0.2499 & 0.5295 & -0.2043 & -0.0753 \\
\textbf{Hispanic}& -0.2285 & -0.2089 & 0.5075 & -0.0701 \\
\textbf{Asian} & -0.1776 & -0.2178 & -0.1924 & 0.5879 \\
\bottomrule
\end{tabular}
\end{center}

\vspace{0.2cm}

\begin{center}
\captionof{table}{ Gravity-Based Segregation Matrix ($M_{gs}$) for February \\ 2020 under the 70\% threshold} \label{tab:tableA15}
\vspace{0.1cm}
\begin{tabular}{@{} l | C C C C @{}}
\toprule
 & \textbf{White} & \textbf{Black} & \textbf{Hispanic} & \textbf{Asian} \\
\midrule
\textbf{White} & 0.2077 & -0.1465 & -0.0523 & -0.0088 \\
\textbf{Black} & -0.0636 & 0.1046 & -0.0327 & -0.0084 \\
\textbf{Hispanic}& -0.0199 & -0.0778 & 0.1076 & -0.0100 \\
\textbf{Asian} & -0.0867 & -0.1745 & -0.0969 & 0.3580 \\
\bottomrule
\end{tabular}
\end{center}

\setcounter{figure}{2}
\begin{figure}[H]
    \centering 
    \includegraphics[width=17 cm,trim=0 0 0 10,clip]{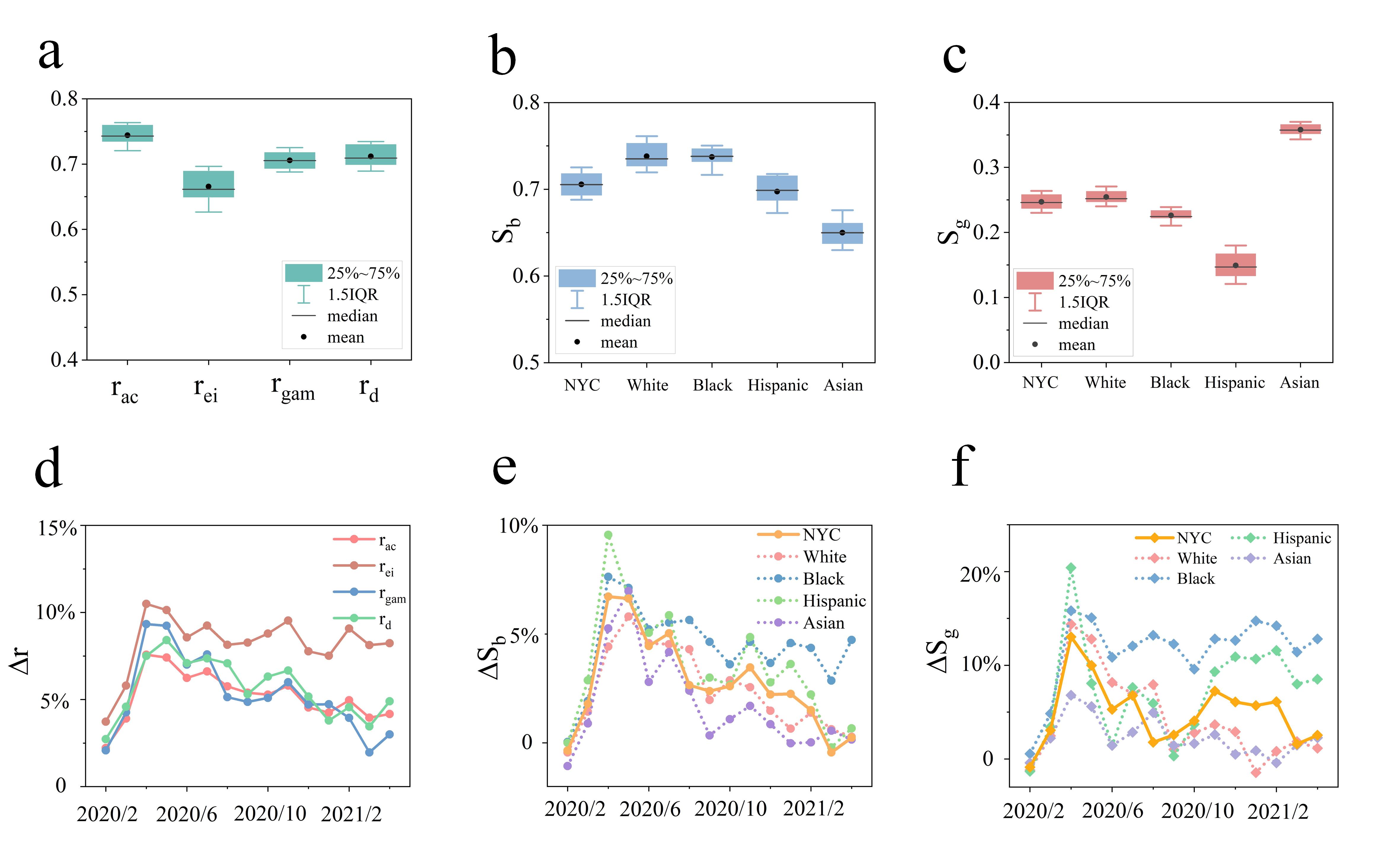}
    \caption{\textbf{a-c} Boxplots of assortativity indices ($r$), baseline segregation indices ($S_b^i$) and gravity-based segregation indices ($S_g^i$) across racial groups during the 14-month pre-pandemic period under the $70\%$ threshold, with all values significantly greater than the values under the $50\%$ baseline 
\textbf{d-f} Monthly fluctuations in $r$, $S_b$ and $S_g$ during the pandemic under the $70\%$ threshold reveal intensified racial isolation during the pandemic. All indices show significant increase after the implementation of the control measures.}
	\label{fig:fig_si_70_seg}
\end{figure}

\begin{figure}[H]
    \centering 
    \includegraphics[width=15 cm,trim=0 0 0 10,clip]{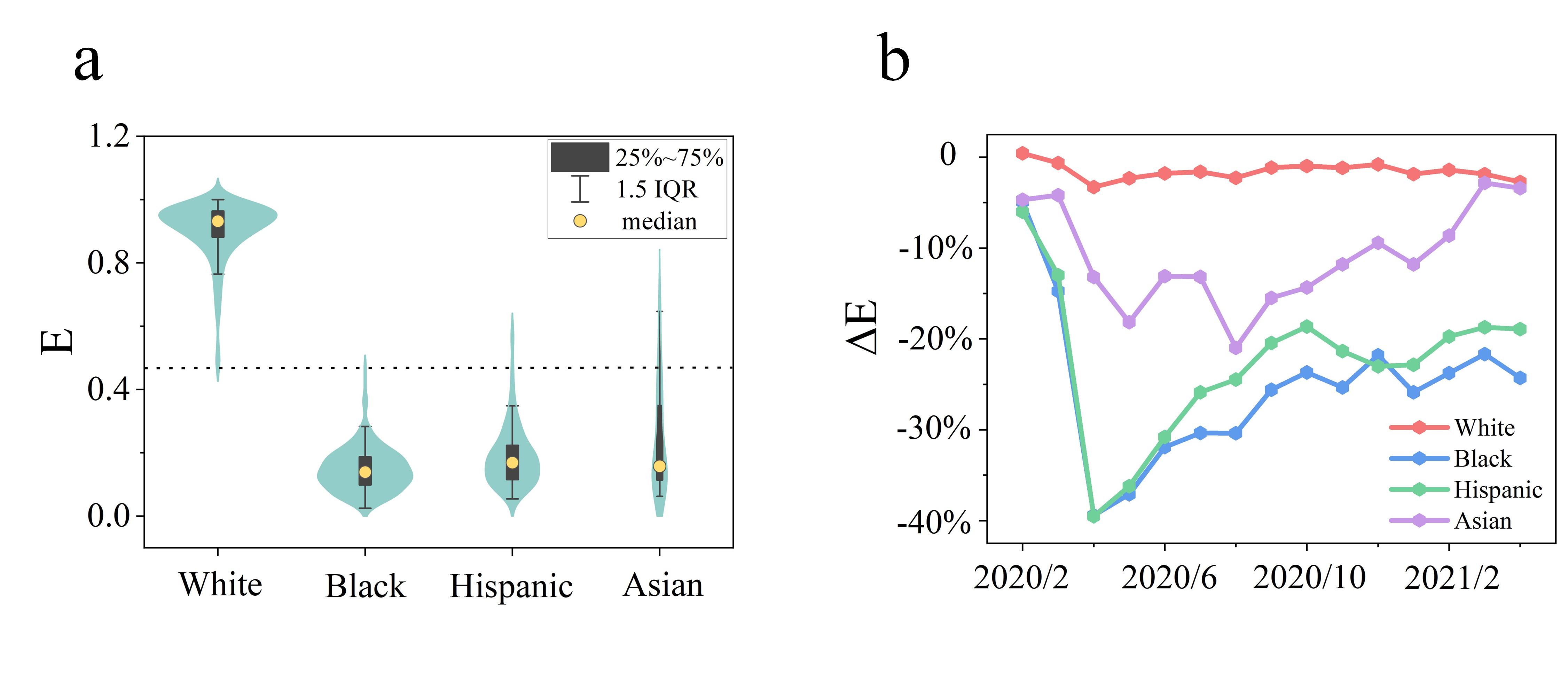}
    \caption{
\textbf{a} Violin plots of exposure rates by racial group under the $70\%$ threshold (February 2020). White-majority CTs still present higher mean exposure and broader distribution compared to Black, Hispanic, and Asian-majority CTs.
\textbf{b} Percent change in average exposure rates relative to pre-pandemic baseline (14-month mean) across racial groups over time under the $70\%$ threshold. White-majority CTs exhibited the smallest decline, while Black, Hispanic, and Asian-majority CTs displayed larger reductions.}
	\label{fig:fig_si_exposed_70}
\end{figure}

\begin{figure}[ht!]
    \centering 
    \includegraphics[width=13cm,trim=0 0 0 10,clip]{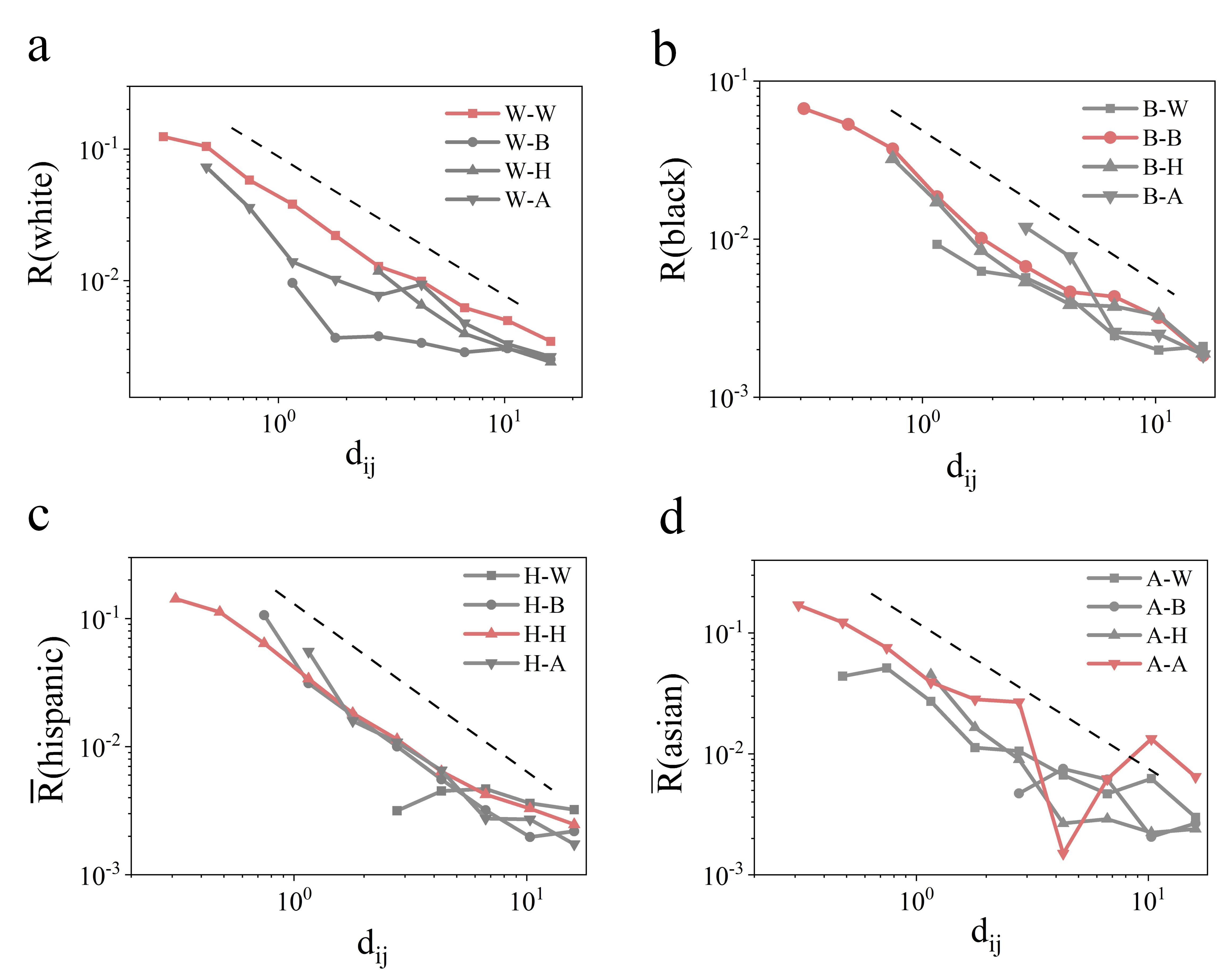}
    \caption{\textbf{a-d} Average flow proportion $\bar R$ decays with $d_{ij}$ with a slope $\bar R \propto d_{ij}^{\eta}$ under the $70\%$ threshold (February 2020), in which $\eta \in[-1.2,-1.1]$ (White: W; Black: B; Hispanic: H, Asian: A). Curves representing flows between the same race CTs maintain their dominance across most distance scales.}
	\label{fig:si_distance_70}
\end{figure}

\clearpage

\begin{table}[h!]
  \centering
  \caption{Gravity Model Specifications under the 70\% threshold \\(November 2019 - February 2020)}
  \label{tab:model_2019_70}
  \setlength{\tabcolsep}{25pt} 
  \begin{tabular}{lcc}
     \toprule
      \textbf{Parameter} & \textbf{Baseline (w/o $h_{ij}$)} & \textbf{Full (w/ $h_{ij}$)} \\
  \midrule
    \multicolumn{3}{l}{\textbf{Conventional Parameters}} \\
    $\lambda$ & 2.814***  & 0.597***   \\
              & (0.044)   & (0.014)    \\
    $\alpha$  & 2.647***  & 2.370***   \\
              & (0.008)   & (0.007)    \\
    $\beta$   & 1.472***  & 1.404***   \\
              & (0.005)   & (0.005)    \\
    $\delta$  & 1.267***  & 1.157***   \\
              & (0.004)   & (0.003)    \\
    \midrule
    \multicolumn{3}{l}{\textbf{Racial Parameter}} \\
    $\gamma$  & \multicolumn{1}{c}{--} & 3.401*** \\
              &                       & (0.033)  \\
    \midrule
    Observations & \multicolumn{2}{c}{138,484} \\
    $R^2$       & 0.558  & 0.616  \\
    Adjusted $R^2$ & 0.558 & 0.616 \\
    \bottomrule
     \multicolumn{3}{l}{\footnotesize \textit{Notes: *$p < 0.1$; **$p < 0.01$; ***$p < 0.001$.}}
  \end{tabular}
\end{table}

\vspace{1.5cm} 

\begin{table}[h!]
  \centering
  \caption{Gravity Model Specifications under the 70\% threshold \\(March 2020 - June 2020)}
  \label{tab:model_2020_70}
  \setlength{\tabcolsep}{25pt} 
  \begin{tabular}{lcc}
     \toprule
      \textbf{Parameter} & \textbf{Baseline (w/o $h_{ij}$)} & \textbf{Full (w/ $h_{ij}$)} \\
  \midrule
    \multicolumn{3}{l}{\textbf{Conventional Parameters}} \\
    $\lambda$ & 36.013*** & 2.225***   \\
              & (0.458)   & (0.065)    \\
    $\alpha$  & 2.027***  & 1.991***   \\
              & (0.010)   & (0.009)    \\
    $\beta$   & 1.293***  & 1.311***   \\
              & (0.007)   & (0.006)    \\
    $\delta$  & 1.161***  & 1.032***   \\
              & (0.006)   & (0.005)    \\
    \midrule
    \multicolumn{3}{l}{\textbf{Racial Parameter}} \\
    $\gamma$  & \multicolumn{1}{c}{--} & 4.606*** \\
              &                       & (0.042)  \\
    \midrule
    Observations & \multicolumn{2}{c}{107,216} \\
    $R^2$       & 0.428  & 0.516  \\
    Adjusted $R^2$ & 0.428 & 0.516 \\
    \bottomrule
     \multicolumn{3}{l}{\footnotesize \textit{Notes: *$p < 0.1$; **$p < 0.01$; ***$p < 0.001$.}}
  \end{tabular}
\end{table}


\end{appendices}
\end{document}